# New Approach to the Characterization of $M_{max}$ and of the Tail of the Distribution of Earthquake Magnitudes

V.F. Pisarenko[1], A. Sornette[2], D. Sornette[3,4] and M.V. Rodkin[5]

[1]International Institute of Earthquake Prediction Theory
and Mathematical Geophysics
Russian Ac. Sci. Warshavskoye sh., 79, kor. 2, Moscow 113556, Russia

[2]ETH Zurich, Swiss Seismological Service
HPP P6.1, Hönggerberg, CH-8093 Zürich, Switzerland

[3]ETH Zurich, D-MTEC, Kreuzplatz 5
CH-8032 Zurich, Switzerland

[4]Institute of Geophysics and Planetary Physics
and Department of Earth and Space Science
University of California, Los Angeles, California 90095

[5]Geophysical Center of
Russian Ac. Sci. Molodezhnaya 3, Moscow 119206, Russia

**Abstract:** We develop a new method for the statistical estimation of the tail of the distribution of earthquake sizes recorded in the Harvard catalog of seismic moments converted to $m_W$-magnitudes (*1977-2004* and *1977-2006*). For this, we suggest a new parametric model for the distribution of main shock magnitudes, which is composed of two branches, the pure Gutenberg-Richter distribution up to an upper magnitude threshold $m_1$, followed by another branch with a maximum upper magnitude bound $M_{max}$, which we refer to as the two-branch model. We find that the number of main events in the catalog ($N = 3975$ for *1977-2004* and $N=4193$ for *1977-2006*) is insufficient for a direct estimation of the parameters of this model, due to the inherent instability of the estimation problem. This problem is likely to be the same for any other two-branch model. This inherent limitation can be explained by the fact that only a small fraction of the empirical data populates the second branch. We then show that using the set of maximum magnitudes (the set of *T*-maxima) in windows of duration *T* days provides a significant improvement, in particular (i) by minimizing the negative impact of time-clustering of foreshock / main shock / aftershock sequences in the estimation of the tail of magnitude distribution, and (ii) by providing via a simulation method reliable estimates of the biases in the Moment estimation procedure (which turns out to be more efficient than the Maximum Likelihood estimation). We propose a method for the determination of the optimal choice of the *T*-value minimizing the Mean Square Error of the estimation of the form parameter of the GEV distribution approximating the sample distribution of *T*-maxima, which yields $T_{optimal}=500$ days. We have estimated the following quantiles of the distribution of *T*-maxima for the whole period 1977-2006: $Q_{16\%}(M_{max})= 9.3$, $Q_{50\%}(M_{max})= 9.7$ and $Q_{84\%}(M_{max}) = 10.3$. Finally, we suggest two more stable statistical characteristics of the tail of the distribution of earthquake magnitudes: the quantile $Q_T(q)$ of a high probability level *q* for the *T*-maxima, and the probability of exceedence of a high threshold magnitude $\rho_T(m^*) = P\{ m_k \geq m^*\}$. We obtained the following sample estimates for the global Harvard catalog $\hat{Q}_T (q=0.98) = 8.6 \pm 0.2$ and $\hat{\rho}_T (8) = 0.13\text{-}0.20$. The comparison between our estimates for the two periods *1977-2004* and *1977-2006*, where the later period included the great Sumatra earthquake *24.12.2004, $m_W$=9.0* confirms the instability of the estimation of the parameter $M_{max}$ and the stability of $Q_T(q)$ and $\rho_T(m^*) = P\{ m_k \geq m^*\}$.

## 1. Introduction

In a series of papers, Gutenberg and Richter *(1942, 1954, 1956)* suggested their famous formula

$$log \ N_T(m) = a_T - b \ m, \tag{1}$$

with parameters $a_T$ and $b$. The formula (1) gives the number of earthquakes $N_T(m)$ with magnitudes $M \geq m$, occurring in a large seismic zone or at the global scale for a sufficiently long time period $T$. Relation (1), referred to as the *magnitude-frequency law,* has become one of the fundamental laws of seismology. With the assumption that seismicity is stationary, the Gutenberg-Richter law describes the frequency $\lambda(m)$ for the occurrences of earthquakes with magnitudes exceeding $m$,

$$log \ \lambda(m) = \alpha - \beta \ m, \ \text{or equivalently} \ \lambda(m) = 10^{\alpha - \beta m}, \tag{2}$$

where

$$\lambda(m) = \lim_{T \to \infty} N_T(m) \ / \ T. \tag{3}$$

The parameter $\alpha$ characterizes the seismicity level in the particular zone of study, while the parameter $\beta$ (defined as the slope of the magnitude frequency law) quantifies the relative frequency of large versus small earthquakes.

More generally, if one fixes a lower threshold $m_0$ of registered magnitudes then, after normalization of eq. (2) by $\lambda(m_0)$, the frequency $\lambda(m)$ can be written

$$\lambda(m) \ / \ \lambda(m_0) \ = \ \overline{F} \ (m), \qquad m \geq m_0, \tag{4}$$

where $\overline{F} \ (m)$ is some non-increasing function, varying from 1 down to 0. In the case for which the Gutenberg-Richter law holds, we have:

$$\overline{F} \ (m) = 10^{-\beta(m - m_0)}. \tag{5}$$

This implies that the function $F(m) = 1 - \overline{F} \ (m)$, defined as the complementary of $\overline{F} \ (m)$ with respect to unity, can be considered as a distribution function of earthquake sizes. Note that expression (4) does not contain any information on the seismicity level, which would be necessary for assessing seismic risks. It is only intended to describe the characteristics of the distribution of earthquake sizes. The exponential law (5) for the complementary distribution of magnitudes (5) corresponds to a power law distribution of seismic moments.

If we apply, as it is often done, formulae (1)-(5) to the *main shocks* of some catalog, and if we assume that the temporal flow of events is *Poissonian* then, for a fixed number of events $n$, the realized magnitudes $m_1, ..., m_n$ can be considered as a sample of iid rv (independent identically distributed random variables) with distribution function $F(m)$. In the case of a Poissonian flow of main shocks, the relation between the temporal flow of earthquakes and a sample of $n$ identically distributed independent random values is established. If the time span covered by the catalog grows, the random (Poissonian) number $n$ tends with probability one to infinity, and asymptotic statistical inference obtained for the sample of the $n$ iid rv become valid for Poissonian flow of events. Though, statistical inferences based on the condition of a fixed finite $n$ is of interest by itself too.

The Gutenberg-Richter (G-R) law has been subjected to numerous investigations (see, e.g. Bird and Kagan, 2004; Cosentino et al., 1977; Kagan, 1991; 1996; 1999; 2002a; 2002b; Kijko and Sellevol 1989, 1992; Knopoff et al., 1982; Main et al., 1999; Ogata and Katsura, 1993; Pisarenko and Sornette, 2003; 2004; Sornette et al., 1996; Utsu, 1999; Wu, 2000). One can summarize the present situation by saying that, for small and moderate magnitudes, and for large space-time volumes, the Gutenberg-Richter law is valid with a large degree of accuracy. However, for the largest magnitudes, some more or less significant deviations from (1) have been documented (see e.g. Pisarenko and Sornette (2004) and references therein). Their investigation is hampered by the insufficient number of large earthquakes. Inevitably, existing models of the deviations from the G-R and the numerous proposals to modify it for large magnitudes suffer from large statistical uncertainty. As a consequence, the problem of finding an adequate description of the tail of the magnitude distribution cannot be considered as definitely settled. One of the best known modifications of the G-R (Kagan, 1997; Kagan and Schoenberg, 2001) consists in multiplying the power law distribution of seismic moments (which, as we recalled above, corresponds to the G-R exponential distribution of magnitudes) by an exponential factor (also referred to as an exponential taper) which leads to a Gamma-distribution for seismic moments. The characteristic moment in the exponential taper is often referred to as the "corner" moment, as it constitutes the typical magnitude at which the distribution departs significantly downward from the pure G-R law. The corner moment is not the absolute maximum size and larger earthquake moment are authorized in this model, albeit with exponentially smaller probability. This exponential taper to the power law of seismic moments constitutes one among several attempts to take into account the presence of a downward curvature observed in the empirical distribution of earthquake sizes for the largest moments (or magnitudes). See Bird and Kagan (2004) for the use of the "Kagan" model to determine the "corner" magnitude for seven different tectonic settings.

Rather than introducing a "soft" truncation of the G-R law, a different class of models assume that the G-R law holds up to a maximum magnitude $M$, beyond which no earthquake are observed (Cosentino et al., 1977; Dargahi-Noubary, 1983; Main et al., 1999; Pisarenko, 1991; Pisarenko et al., 1996)

$$F(x) = \begin{cases} 0; & x < m ; \\ [\, 10^{-\beta m} - 10^{-\beta x}\,]\,/\,[\, 10^{-\beta m} - 10^{-\beta M}\,]; & m \leq x \leq M ; \\ 1; & x > M . \end{cases} \quad (6)$$

The parameter M represents the *maximum possible earthquake size: $M = M_{max}$*. This parameter plays a very important role in seismic risk assessment and in seismic hazard mitigation (see e.g. Bender and Perkins, 1993; Pisarenko et al., 1996; Kijko and Graham, 1998; Kijko et al., 2001). It should be noted that the truncated G-R (6) ensures the finiteness of the mean seismic energy, whereas the G-R in its unlimited form (1)-(5) corresponds to a regime with infinite seismic energy, which is, of course, an undesirable property of the model (the exponential taper in the Kagan model of seismic moment distribution ensures also the finiteness of the mean seismic moment (Knopoff and Kagan, 1977)). The parameter $M_{max}$ is very suitable for engineers and insurers: having a reliable estimate of $M_{max}$, it is comparatively easy to take adequate decisions on the construction standards of buildings or about insurance policy. As a consequence, the modified G-R (6) has undergone a wide dissemination. Unfortunately, all attempts so far for a reliable statistical estimation of $M_{max}$ did not give satisfactory results. The statistical scatter of its estimates and their reliability are far from the desirable level. Attempts to attribute a maximum magnitude to individual faults rather than to regions suffer from the same problems and in addition face the fact that many large earthquakes involve several faults or fault segments which are difficult if not impossible to determine in advance (Black et al., 2004; Ward, 1997).

Below, we explain and illustrate the intrinsically unstable properties of the determination of $M_{max}$. Besides, $M_{max}$ has the following undesirable features.

1. $M_{max}$ is ill-defined. It does not contain **the time interval (or, time scale)** over which it is valid. Suppose we are supplied with a very good estimate of $M_{max}$. Is this value related to the whole geologic history (4.5*10^9 years), or to the last geological time period since the Gondwana accretion (200*10^6 years), or to the last 200 years? This question, we believe, should be very important for practical applications, but $M_{max}$ does not answer it at all. A way to circumvent this problem is to define $M_{max}$ as the magnitude of an exceedingly low probability event over a finite time period, say the next 30 years. This last definition is more in line with the needs of insurance and risk policy.

2. It is a fact that the maximum earthquake magnitude should be finite because the Earth is finite. But using the "small increment" argument (where $\varepsilon$ is an arbitrary small positive number), one can always argue that "if the magnitude, say $M^*$, is the maximum possible, then why is magnitude $(M^* + \varepsilon)$ impossible? It is difficult to answer this question properly.

3. Finally, the parameter $M_{max}$ is highly unstable from a statistical point of view. We are going to demonstrate this fact below.

Thus, in spite of its practical attractiveness and its implication for constraining physical models and for risk assessment, we believe that $M_{max}$ cannot be used fruitfully in applications. The purpose of the present paper to re-examine the issue of the tail of the distribution of large earthquake magnitude by suggesting other characteristics of the tail behavior that are more stable that $M_{max}$.

The paper is structured as follows. Section 2 presents the general background on the theory of extreme value statistics and applies it to the set of $T$-maxima, the maximum magnitudes in time windows of duration equal to $T$ days. Section 2.1 and Appendix A describe the declustering algorithm used here to prepare a set of $N$ main shocks which are further analyzed in the remaining of the paper. Section 2.2 and Appendix B presents the main results on the estimation of the parameters of the Generalized Extreme Value (GEV) distribution fitted to the empirical distribution of $T$-maxima of the global Harvard catalogues of shallow earthquakes. Section 2.3 stresses the inherent instability of the determination of $M_{max}$ and suggests two more stable characteristics of the tail behavior. Section 2.4 presents a reshuffling procedure that improves the estimates and provide in addition confidence intervals. Section 3 and Appendix C presents the two-branch model which is at the basis of our simulation procedure. The later allows us to quantify rather precisely the strong biases in the estimation of the parameters of the GEV distribution. Section 3.1 presents the method to determine the optimal time window $T$, subsequently used in the determination of $M_{max}$ and of the other tail characteristics. Section 3.2 derives our main results and presents the skewed distributions of the different tail characteristics. Section 4 summarizes all our results and gather together all the main numerical estimations of the tail characteristics.

## 2. Estimation of $M_{max}$ in $T$-windows using the Generalized Extreme Value Theory applied to the declustered catalogs

### 2.1 Preparation of the data

Our different analyses presented below are performed on a declustered catalog, in which aftershocks have been removed more or less exhaustively according to a standard method described and tested in Appendix A. As shown in Appendix A, the stationarity of the declustered catalog of main shocks can be considered satisfactory for the time period 01.01.77-20.12.04 over the magnitude range $m_W \geq 5.5$. Two standard statistical tests do not reject the hypothesis that these main shocks obey a Poisson process with a constant rate.

Let us consider such a declustered catalog. It is a sample $x_1, \ldots x_N$ of $N$ iid rv (independently identically distributed random variables) representing the magnitude measurements of main shocks in a catalog that occurred at the corresponding times $t_1, \ldots t_N$. Following Epstein and Lomnitz (1966), let us then divide the time interval $(t_1, t_N)$ covered by the catalog into $n$ segments of length $T$ (days) ($n \cong (t_N - t_1)/T$). Within each time segment, we measure the maximum magnitude that we denote as $m_k$. In this way, we get a sample of $n$ maxima $m_1, ..., m_n$. While this reduces the set of events from $N$ to $n$, it turns out that the distribution of these $T$-maxima is more appropriate for estimating the tail of the distribution of earthquake magnitudes. Indeed, the construction of the sample of $T$-maxima from the catalog of main shocks amounts in some sense to applying a "large magnitude" filter which emphasizes large magnitudes and removes lower magnitudes. There are several advantages in applying this filter.

First, we minimize the influence of the a priori unknown random temporal process governing the time occurrence $t_1, ... t_N$ of the $N$ events, which allows us to focus exclusively on the distribution of magnitudes. Fixing a time interval T and measuring the maximum magnitude $m_{max}(T)$ within this interval avoids the problem associated with the random number of events. As a result of such operation, we obtain the sequence of iid rv $m_{max}(T_1)$, $m_{max}(T_2)$, ... , whereas, the initial declustered catalog is a sample of a marked point process $m_1, m_2, ...$ that occurred at random times $t_1, t_2, ...$ This has a significant impact, for instance in writing the likelihood function. Furthermore, this construction of $n$ $T$-maxima significantly weakens the disturbances coming from remaining "aftershocks" and other triggered events, that may not have been fully removed by the declustering method described in Appendix A. Indeed, aftershocks have more often lower magnitudes and thus enter in the sequence of $m_{max}(T_1)$, $m_{max}(T_2)$, ... with small probability.

The study of the rv $m_{max}(T_1)$, $m_{max}(T_2)$, ... , $m_{max}(T_n)$ is more convenient that of $M_{max}$. While the later is ill-defined as we mentioned above, the *maximum magnitude $m_k$ of events that occurred in the time interval (0; T) is well-defined* and perhaps more fruitful for practical users. Note that if the DF of single event magnitudes is $F(x)$, and if the sequence of events is a Poissonian sequence with intensity $\lambda$, then the DF for the maximum magnitude $m_k$ in a $T$-window is denoted as $\Phi_T(x)$ and is given by

$$\Phi_T(x) = P\{ m_k < x; \ n \geq 1 \} = \sum_{k=1}^{\infty} F^k(x) \ exp(-\lambda T) \ (\lambda T)^k/k! \ / \ [1 - exp(-\lambda T)] =$$

$$= \{ exp[-\lambda T(1 - F(x))] - exp(-\lambda T) \} \ / \ [1 - exp(-\lambda T)]. \qquad (7)$$

When one can assume that the initial sample can be safely modeled by iid rv with some magnitude distribution, then using the full data set rather than the reduced set of n T-Maxima is superior since, in principle, the whole initial sample is generally more informative than any reduced sample (Knopoff and Kagan, 1977). But this requires a rather strong faith in the ability of the declustering algorithm so that the main shock time occurrences are truly Poissonian. Otherwise, the distribution of time occurrences inevitably enters into the likelihood of the sample. While the two tests in Appendix A cannot reject this hypothesis for the declustered catalog, there are still remaining dependencies between the main shocks. These dependencies are strongly reduced by working with the set of $n$ $T$-maxima. Using the $n$ $T$-maxima $m_1, ..., m_n$ significantly weakens the influence of aftershocks and the possibly complex time dependence between events, and makes possible the application of general results of the theory of extreme values (of course under the assumption that the initial distribution of magnitudes satisfies the conditions of the theorems of this theory).

## 2.2 *Estimation of the three parameters (m, s, ξ) of the GEV distribution of $T$-maxima $m_k$*

We thus assume that the initial distribution of magnitudes satisfies the conditions of extreme value theory. Consequently, the distribution of the $T$-maxima $m_k$ can be satisfactorily approximated by the limit distribution of extreme value theory, named the Generalized Extreme Value (GEV) distribution. The GEV distribution function (DF) depends on 3 parameters *(m, s, ξ)* and is given as:

$$F(x) = exp\{ -[1+(\xi/s)*(x-m)]^{-1/\xi} \}; \quad 1+(\xi/s)*(x-m)] > 0; \quad \xi \neq 0; \tag{8}$$

$$F(x) = exp\{ -exp[-(x-m)/s] \}; \qquad \xi = 0, \tag{9}$$

where $m$ is the location parameter; $s$ is the scale parameter, and $\xi$ is the form parameter. In accordance with eq.(8), the domain of non-zero probability depends on the parameters, which makes the GEV family non-regular. In other words, this family does not satisfy the conditions of regularity required by the standard theory of maximum likelihood estimation.

Our central point consists in assuming that the maximum magnitude is bounded. Thus, in accordance with eq.(8), only **negative** values of the form parameter $\xi$ are acceptable, which give the following upper limit $M_{max}$ for magnitudes:

$$m_k \leq M_{max} = m - s/\xi ; \quad \xi < 0. \tag{10}$$

Because the number $n$ of $T$-maxima is not large, it is important to ask if the Maximum Likelihood Estimates (*MLE*) is the most efficient. Appendix B compares three well-known methods for the statistical estimation of GEV-parameters ($m, s, \xi$): the Maximum Likelihood Estimates (*MLE*), the Moment Estimates (*ME*) and estimates based on Probability-Weighted Moments. For small and moderate sample sizes ($n \cong 10 \div 50$), the *MLE* are less efficient than the *ME* and the *PWM* estimate, whereas in turn the *ME* is slightly better than the *PWM* estimate. These differences disappear for larger sample size ($n = 200$). Thus, we shall use the Moment Estimates as the most efficient for estimation of GEV-parameters in our situation. We choose the window size $T$ as a compromise between two conflicting requirements. While a smaller $T$ increases the sample size $n$, $T$ can not be too small, since the limit theory of extreme values requires a large number of observations in each $T$-interval. With too small $T$'s, we can even get empty $T$-intervals with no event. In contrast, $T$ cannot be very large, since $n$ would be small, and the estimation of parameters would be inefficient, as shown in Appendix B. How can we judge on these conflicting conditions? The first condition demands that we should not have (or, almost not have) empty intervals. In our case, this condition demands $T \cong 15$ days, or larger. For $T=10$, the percentage of non-empty $T$-intervals is *98.04%;* for $T=15$ percent, it is *99.71*, and only from $T=16$ and larger, we have *100%* of non-empty $T$-intervals. On the other hand, the scatter of ME-estimates makes us take $T$ not more than, say, *300* since our time span (*10214 days*) is such that for $T = 300$ we have only $n \cong 30$ which we take as the lower limit for a reliable estimation of the three parameters ($m, s, \xi$).

We thus estimate the three parameters ($m, s, \xi$) for the sample $m_1 ,..., m_n$ by the method of Moments described in Appendix B. We also performed all the estimations presented below with the Maximum Likelihood method and confirm the known fact that the method of Moments is more efficient for small sample size $n$ of the order of *50* or less (see *Hosking et al. 1985, Coles and Dixon 1999*). Therefore, we only report our results obtained with the method of moments. We present our results for the Harvard catalog of seismic moments in the time interval *01.01.1977-18.12.2004*. This interval was chosen because it does not contain the largest event on 24 December 2004 and some of its foreshocks. Very similar results are obtained for the subset of shallow earthquakes confined to subduction zones, confirming the stability of our estimates.

Fig.1 shows the Moment Estimation (ME) of the parameter $\xi$ for the global Harvard catalog of shallow events (*01.01.77-18.12.04; h $\leq$ 70 km; $M_w \geq$ 5.5; N = 3975 main shocks*) as a function of the time duration $T$. It is clear from the figure that the accuracy of the $\xi$-estimate depends on $T$. Similar dependences occur for the other parameters, although their dependence is not as dramatic. We show these dependences in Table 1 for a set of $T$-values.

Table 1. Moment-estimates of the parameters of the GEV distribution of the *T*-maxima for the global Harvard catalog of shallow events (*01.01.77-18.12.04; h ≤ 70 km; M_w ≥ 5.5; N = 3975 main shocks* ).

| T, days | 12 | 30 | 50 | 100 | 200 |
|---|---|---|---|---|---|
| $\hat{m}$ | 6.23 | 6.65 | 6.90 | 7.24 | 7.52 |
| $\hat{s}$ | 0.46 | 0.49 | 0.47 | 0.41 | 0.35 |
| $\hat{\xi}$ | -0.056 | -0.140 | -0.180 | -0.222 | -0.263 |
| $\hat{M}_{max}$ | 14.50 | 10.15 | 9.49 | 9.08 | 8.86 |

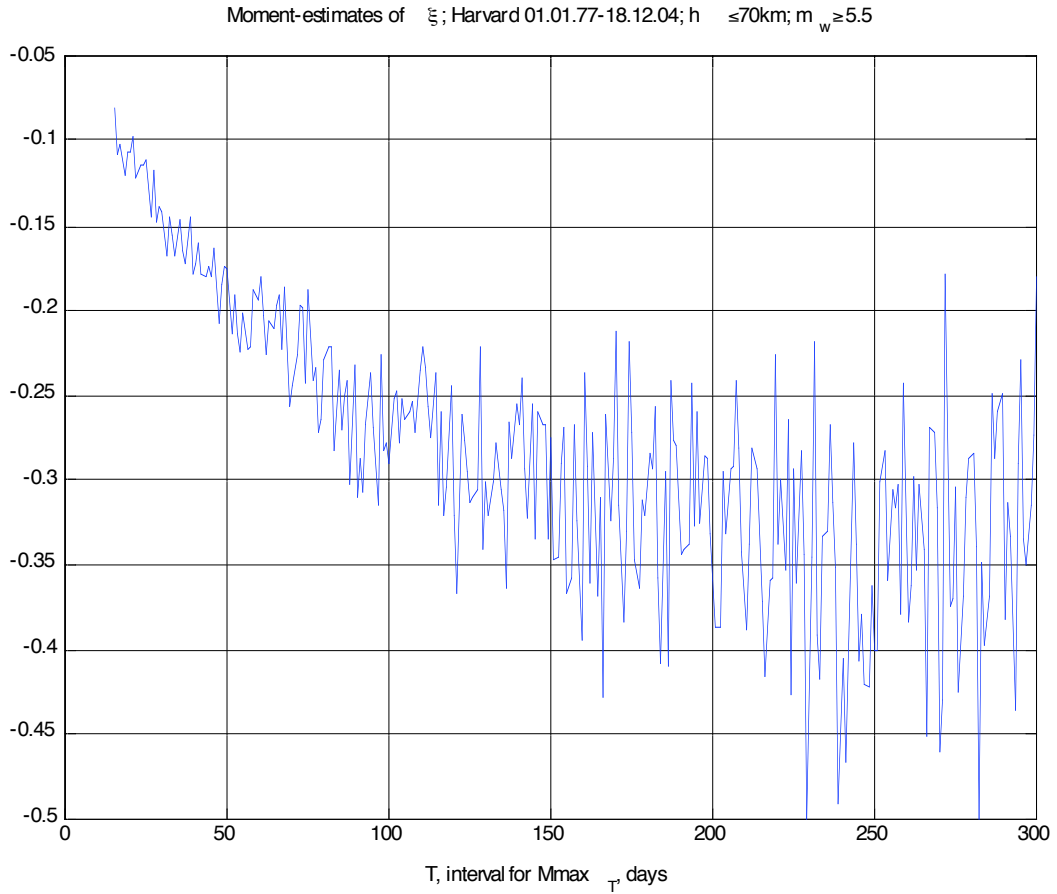

Fig. 1: Moment estimates of the parameter $\xi$ for the global Harvard catalog of shallow events (*01.01.77-18.12.04; h ≤ 70 km; M_w ≥ 5.5; N = 3975 main shocks* ), as a function of the interval duration *T* of the time window in which the *n T*-maxima are determined.

The quality of the fits of the empirical distributions of the *T*-maximum magnitudes $m_k$ by the GEV family can be checked by some goodness-of-fit metric. We have chosen the Kolmogorov distance *KD*:

$$KD = n^{1/2} \, max \mid F(x\mid \hat{m},\hat{s},\hat{\xi}) - \hat{F}(x)\mid, \qquad (11)$$

where $(\hat{m},\hat{s},\hat{\xi})$ are the Moment-estimates of the parameters, *F* is the GEV distribution function (see eq.(8)), and $\hat{F}(x)$ is the sample stepwise distribution function. Since we use a theoretical function with parameters fitted on the data (rather than determined independently of the data), we cannot use the standard Kolmogorov distribution to check the sample value of *KD*. Instead, in order to determine the confidence level of a given *KD*-distance obtained for a given sample, one has to use a numerically simulated distribution of *KD*-distances in the simulation procedure using random GEV samples with the same fitted parameters *($\hat{m}, \hat{s}, \hat{\xi}$ )*. We will show below figures illustrating the range of *T*-values over which the *KD*-distances can be considered to be minimized and stable.

### 2.3 *Instability of $M_{max}$ and better characteristic of the tail behavior*

It is clear from Table 1 that $\hat{M}_{max}$ is quite unstable. The origin of this bad behavior of $\hat{M}_{max}$ can be explained as follows. Consider a sample of limited size *n* (say $n \cong 30$ of one-year-intervals or 120 intervals of length 3 months, as in the Harvard catalog covering the time period 1977-2006), and suppose further that true value of the parameter $\xi_0$ is small in absolute value (-0.1 or smaller which corresponds to the real situation as becomes clear below). Then, it is quite possible that even for the most efficient estimators $\hat{\xi}$ , the random deviations ($\hat{\xi}$ - $\xi_0$) has a standard deviation (std) comparable with the true value $\xi_0$ . In other words, it is quite possible that, due to the random scatter, the estimator $\hat{\xi}$ can take *positive values,* corresponding to unlimited power-like distributions, a result clearly impossible according to our assumption on the bounded nature of the distribution of magnitudes. In such a situation, if we would try to derive the estimator for $M_{max}$ as given by eq.(10), we would get a very unstable estimator, since the parameter ξ is in denominator of expression (10). Thus, the parameter $M_{max}$ is intrinsically unstable in the statistical estimation (under the condition that the studied catalog has not many large earthquakes, which unfortunately is always the case).

We suggest two characteristics that quantify the tail behavior in a better and more stable way:

 1. *The quantile $Q_T(q)$ of the level q of the distribution $\Phi_T(x)$,* i.e. the root of equation

$$\Phi_T(x) = q; \tag{12}$$

 2. *The probability for $m_{max}(T)$ of exceeding some fixed magnitude threshold:*

$$\rho_T(m^*) = P\{ m_k \geq m^*\}, \tag{13}$$

   where *m\* is a fixed (known) rather high threshold.*

Of course, these two characteristics are related with each other, and we consider both only because it could be more convenient to use one of them in a particular application. We are going to show that these characteristics are more stable than $M_{max}$ . The stability of $Q_T(q)$ when the parameter ξ varies in the neighborhood of zero can be deduced from the fact that $Q_T(q)$ unlike $M_{max}$ tends to a finite value as ξ tends to zero, whereas $M_{max}$ tends to infinity in accordance with eq.(10).

Fig.2a shows the quantiles *$Q_T(q=0.98)$* of the GEV distribution of the maximum magnitudes $m_k$ in time intervals of duration *T*, with parameters equal to the ME-estimates obtained from the global Harvard catalog:

$$Q_T(q) = \hat{m} + (\hat{s}/\hat{\xi}) \{[log (1/q)]^{-\hat{\xi}} - 1\}, \quad with \quad q = 98\% . \tag{14}$$

For *T >70 days*, the quantiles *$Q_T(q=0.98)$* stabilize with only a very slow increase as a function of *T*. Even for smaller *T*'s, down to 20 days, *$Q_T(q=0.98)$* is remarkably stable.

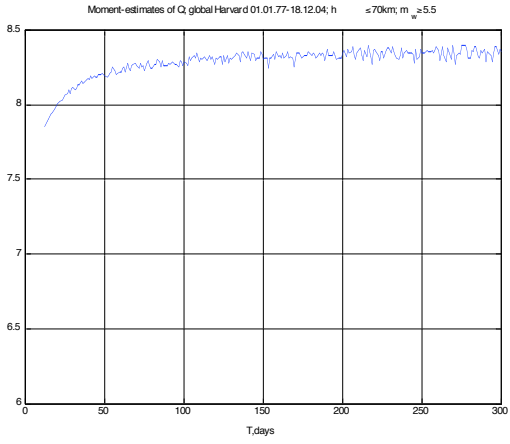
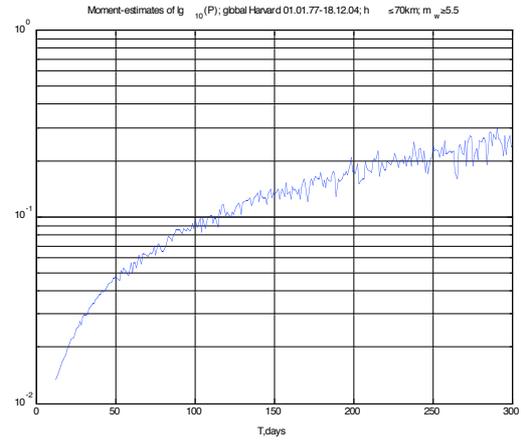

Fig. 2a                    Fig. 2b

Fig. 2a. Quantile $Q_T(q=0.98)$ given by (12) of the GEV distribution of the maximum magnitudes $m_k$ in time intervals of duration $T$ plotted as a function of $T$, for parameters equal to the Moment-estimates obtained from the global Harvard catalog (*01.01.77-18.12.04; h ≤ 70 km; $M_w$ ≥ 5.5; N = 3975 main shocks* ). Fig. 2b shows the Moment-estimates of the probability $P(T)$ defined in (13) that the maximum magnitude in a time window T exceeds *8.0* ($m_k$ ≥ *8.0)*.

Fig. 2b shows the ME-estimates of the probability $P(T)$ that the maximum magnitude in a time window T exceeds *8.0* ($m_k$ ≥ *8.0*). Here again, one can observe the more stable behavior of $P(T)$ compared with that of $M_{max}$. However, unfortunately, in the range $T = 50\text{-}200$, a steady positive increase of $P(T)$ occurs that slows down only above $T \cong 200$. In addition to the more stable behavior of $Q_T(q=0.98)$ and $P(T)$ with respect to variations of $T$, we will see below that these characteristics are as well more stable with respect to random fluctuations.

### 2.4 *Reshuffling procedure and improved ensemble estimates*

Fig.1 shows that there is a strong random scatter of the ξ-estimates for $T \cong 170$ and larger. The scatter looks like a white noise on a background with a smooth trend, thus associated with two correlation components: delta-like and low frequency. It is possible to suppress to some extent the delta component, i.e., the white noise contribution, while the smooth trend cannot be removed without more information (see below). In order to remove the delta-like noise, we use the following method. The initial sample of size $N$ is cut into $n$ segments of length $T$. If $N$ is fixed, then the distribution of occurrence times is uniform in the whole time interval $(t_1, t_N)$ covered by the catalog. But in accordance with our approach, all statistical characteristics would not change if we shuffle the occurrence times randomly over the whole time interval $(t_1, t_N)$. After reshuffling, the sample of $T$-maxima with size $n$ is changed. Therefore, repeating the reshuffling procedure $M$ times, we get $M$ different samples of $T$-maxima, over which we can average. Of course, these different samples are statistically dependent, but nevertheless, additive averaging can reduce the scatter to some extent. What is very important is that we can apply this procedure **to the unique real sample that we have.**

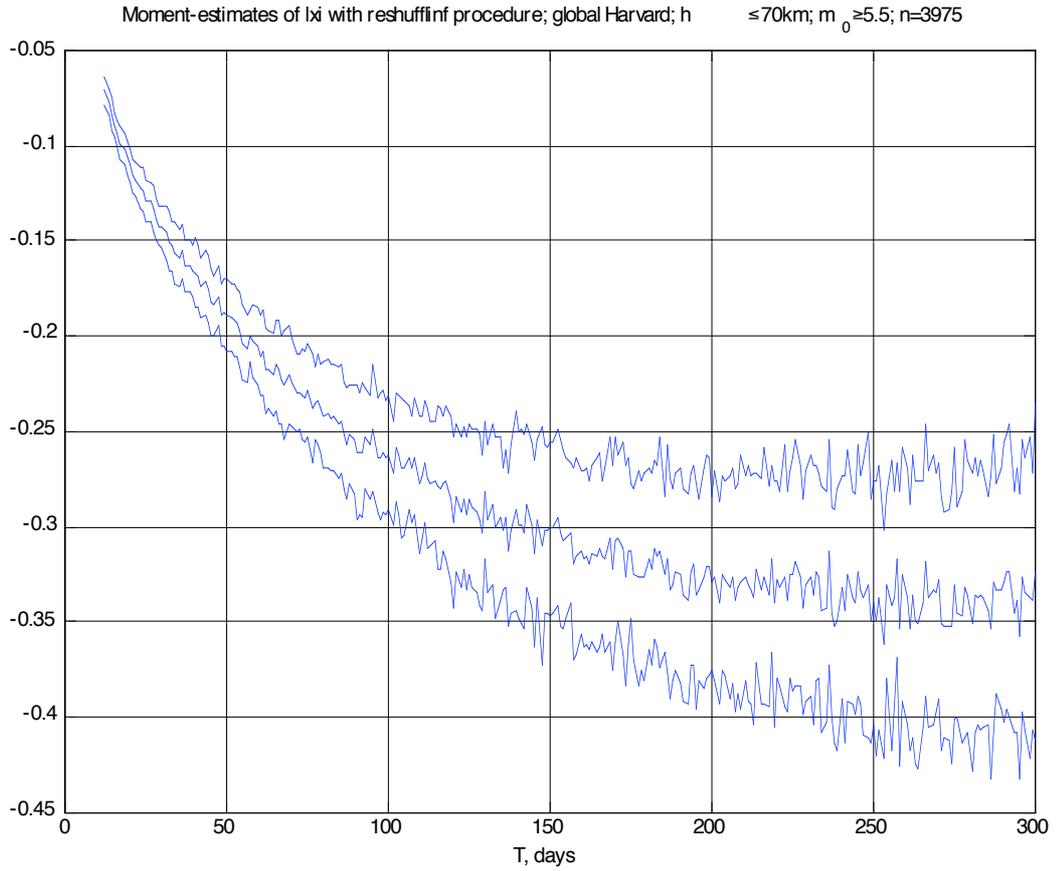

Fig. 3: Moment-estimates of the parameter $\xi$ with the reshuffling procedure as a function of $T$ for the Harvard catalog of shallow events (*01.01.77-18.12.04; h ≤ 70 km; $M_w$ ≥ 5.5; N = 3975 main shocks*). The average (middle curve) as well as the plus-and-minus one standard deviations (top and bottom curves) over 50 shuffled realizations are shown as a function of the interval duration $T$ of the time window in which the *n T*-maxima are determined.

Fig.3 shows the Moment-estimates of the parameter $\xi$ obtained with the reshuffling procedure for the global Harvard catalog of shallow events (*01.01.77-18.12.04; h ≤ 70 km; $M_w$ ≥ 5.5; N = 3975 main shocks*). The average (middle curve) as well as the plus-and-minus one standard deviations (top and bottom curves) over 50 shuffled realizations are shown as a function of the interval duration $T$ of the time window in which the *n T*-maxima are determined. Compared with Fig.1, the high frequency fluctuations have decreased significantly, although the low frequency trend remains. It should be stressed that the standard deviations shown in Fig.3 refer to a single realization, while the averaging over 50 shuffled realizations makes the standard deviation ≅ 7 times smaller, so that our average in Fig.3 (middle curve) has rather low high-frequency noise.

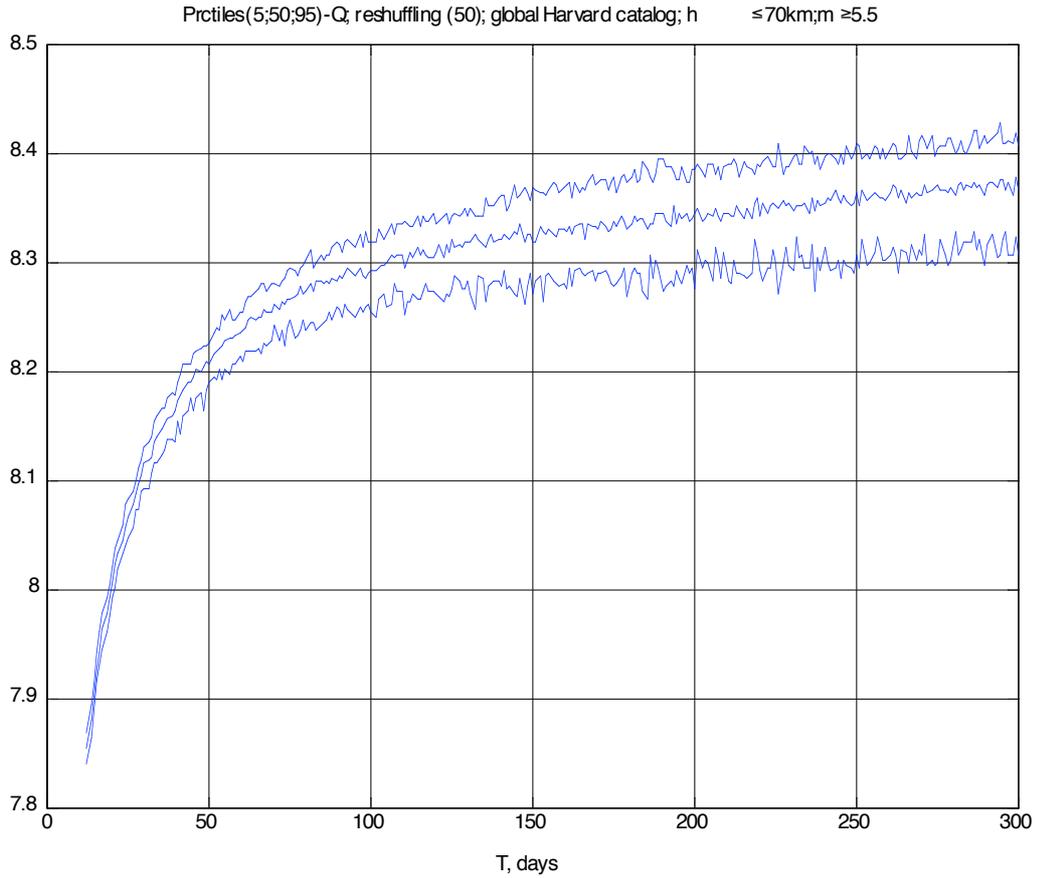

Fig.4: Average (middle curve) as well as the plus-and-minus one standard deviations (top and bottom curves) over 50 shuffled realizations of the ME-estimates for $Q_T(q=0.98)$ as a function of $T$, as obtained with the reshuffling method applied to the Harvard catalog of shallow events (*01.01.77-18.12.04; h ≤ 70 km; M_w ≥ 5.5; N = 3975 main shocks*).

Fig.4 plots the average (middle curve) as well as the plus-and-minus one standard deviations (top and bottom curves) over 50 shuffled realizations of the ME-estimates for $Q_T(q=0.98)$ as a function of $T$, as obtained with the reshuffling method. We see that, for $T > 50$, the $Q$-estimates steadily increase from $Q_T(q=0.98) = 8.33$ up to $Q_T(q=0.98) = 8.39$. It is clear from Fig.4 that that $T$-values less than *50* are inadmissible in the estimation procedure.

Fig. 5 shows the estimation of $M_{max} = m - s/\xi$ obtained with the reshuffling procedure on the Harvard catalog of shallow events (*01.01.77-18.12.04; h ≤ 70 km; M_w ≥ 5.5; N = 3975 main shocks); h ≤ 70 km; M_w ≥ 5.5; N =1287 main shocks*). Since estimates of $M_{max}$ are unstable, we use the order statistics, namely, median, 5% and 95% sample quantiles. The use of the sample mean and of the std would lead to very scattered and unstable estimates. In contrast, the interval between the 5% and 95% sample quantiles can be used as a rather reliable estimate of the 90% confidence interval. We see that the sample median of the 50 shuffled realizations behaves rather smoothly. It monotonically decreases, approaching the maximum observed magnitude in the whole sample (*m_W = 8.4*) as the time window size $T$ increases. This monotonic decrease and the convergence to the maximum observed magnitude is a bad sign of a significant bias of the method.

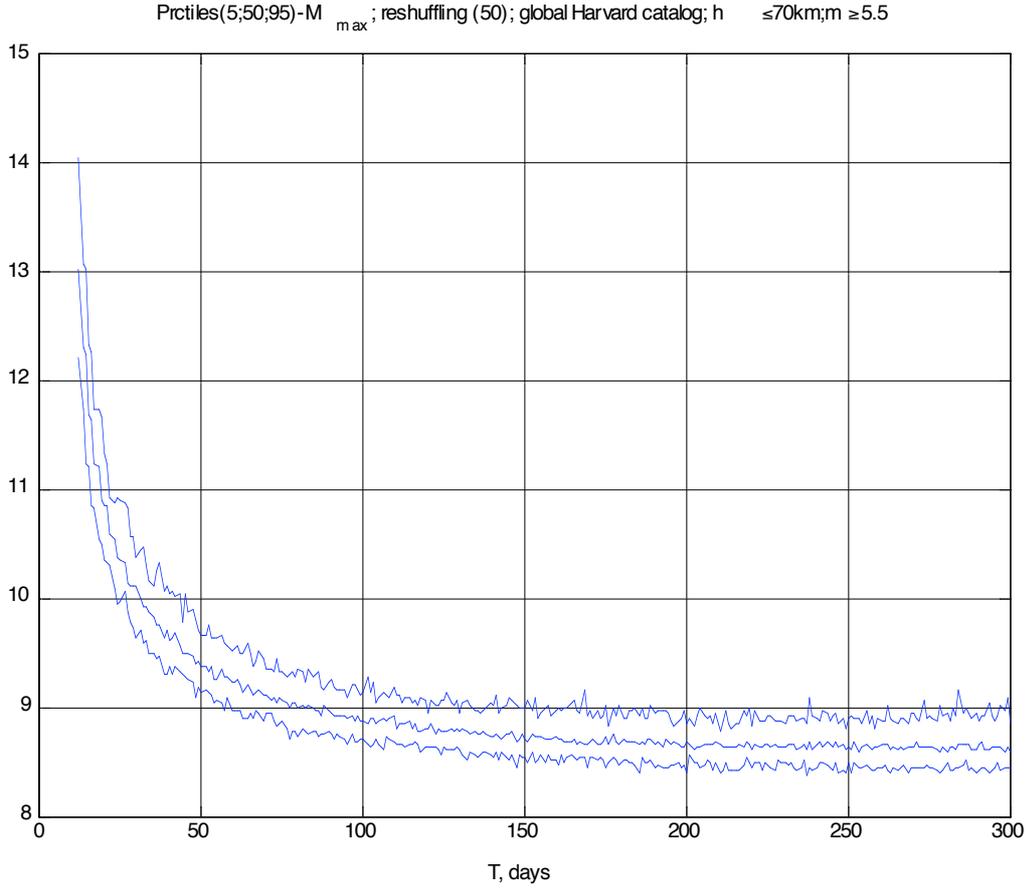

Fig.5: Estimation of $M_{max} = m - s/\xi$ obtained with the reshuffling procedure on the Harvard catalog of shallow events (*01.01.77-18.12.04; h ≤ 70 km; $M_w$ ≥ 5.5; N = 3975 main shocks*); the middle curve corresponds to the average over 50 reshuffled realizations; the top and bottom lines are the r 5% and 95% sample quantiles of these 50 reshuffled realizations.

### 3. Moment Estimation of the GEV-parameters for the Harvard catalog with a simulation procedure minimizing scatter and systematic biases

#### 3.1 Determination of the optimal choice of the size *T* of the windows for the *T*-maxima

Table 1 has made clear that the Moment-estimates of the GEV distribution fitted to the empirical DF of the *T*-maxima depend crucially on the value of *T*. How is possible to choose *T* in an optimal way? We have investigated this problem by simulations using artificial catalogs generated with a realistic two-branch model described in Appendix C. We have applied to these synthetic catalogs the Moment method with reshuffling as explained in section 2.4. We have considered the four sets of the pair of parameters *(m₁, M_max)* for the two-branch model parameterized according to equations (C1-C3):

$$m_1 = 7.5; \ M_{max} = 9.5;$$
$$m_1 = 7.5; \ M_{max} = 10.5;$$
$$m_1 = 8.0; \ M_{max} = 9.5;$$
$$m_1 = 8.0; \ M_{max} = 10.5.$$

(15)

We consider that these values cover possible intermediate values of interest, and we try to determine an optimal value of *T* that would be satisfactory simultaneously for these four sets.

The corresponding values of the parameters *(s,ξ)* can be calculated for each pair as explained in Appendix C.

$m_1$ = 7.5;  $M_{max}$ = 9.5;     $ξ$ = -0.1923; s=0.3846;
$m_1$ = 7.5;  $M_{max}$ = 10.5;    $ξ$ = -0.1370; s=0.4110;
$m_1$ = 8.0;  $M_{max}$ = 9.5;     $ξ$ = -0.2410; s=0.3614;
$m_1$ = 8.0;  $M_{max}$ = 10.5;    $ξ$ = -0.1600; s=0.4000.

The parameter *b* was fixed at the value *b=2.1* (using natural logarithms), which corresponds to *b/ln(10) = 0.912* in the Gutenberg-Richter law. This value corresponds to the best Maximum Likelihood estimate of the slope parameter of the truncated Gutenberg-Richter law for global Harvard sample, *N=3975*, in the magnitude range  ($m_W$ = 5.5 to $m_W$ = 7.8).

Starting from the two-branch model of Appendix C with parameters such as one of the examples in (15), we know that the use of the *T*-maxima must keep the parameters $ξ$ and $M_{max}$ to the values associated with the limit GEV-distribution of corresponding *T*-maxima in the limit where the convergence to the extreme value distribution can be considered to hold. As we discuss in Appendix C, the direct fit of the two-branch model is less efficient that using the *T*-maxima: the distribution of the *T*-maxima is more appropriate for the estimation of the tail of the distribution of magnitudes. The procedure consisting in starting from a two-branch distribution (C1-C3), generating a catalog of *N* main shock from it and then extracting from it a catalog of *n* *T*-maxima is illustrated in Fig.6. This figure shows  (in linear scale) a two-branch density (G-R), the resulting density of *T*-maxima calculated from equation (7), and the approximating GEV density.

When fitting the GEV distribution to the DF of the sample of  *T*-maxima, we have here the luxury of knowing the **exact true** values of the GEV-parameters $ξ$ and $M_{max}$. We can use this knowledge to calculate of **mean square error (MSE)** of the Moment-estimate of the form parameter $ξ$ and parameter  $M_{max}$ . We take the MSE of these parameters for a fixed *T* as a characteristic of the efficiency of the Moment-estimation for this window size *T*:

$$MSE_ξ(T\,|m_1,\,M_{max}\,) =[\,(1/n_s) \sum_{j=1}^{n_s}\, (\,<ξ_T>_j - ξ\,)^2\,]^{1/2}\; ; \qquad (16a)$$

$$MSE_{Mmax}(T\,|m_1,\,M_{max}\,) =[\,(1/n_s) \sum_{j=1}^{n_s}\, (\,<Mmax\,_T>_j - (m - s/ξ)\,)^2\,]^{1/2}\,. \qquad (16b)$$

In expressions (16), $n_s$ *(=25 typically)* is the number of simulations; $<ξ_T>_j$ is the average of the Moment estimates of $ξ$ over $N_S$ *(=50 typically)* shuffles in the j-th simulation;  $ξ$  is the true value of the form parameter. The value of *T* corresponding to the minimum MSE can be considered as **optimal**. Plots of *MSE(T)* as a function of T for the four sets of models (15)  are shown on Fig.7a-h. One can observe that the domain over which the *MSE(T)* remain close to their minima varies from *T=90* up to *T=230*. We can take as a trade-off the value *T =182.5*  (half year) ensuring that the *MSE* is close to its minimal value for all four models (15).

In a first step, we are going to calculate the Moment-estimates of the parameters of the empirical data sets using this *T*-value. In a second step, we want to evaluate the corresponding standard deviation (std) using the information provided by the simulations with the two-branch model (C1-C3).

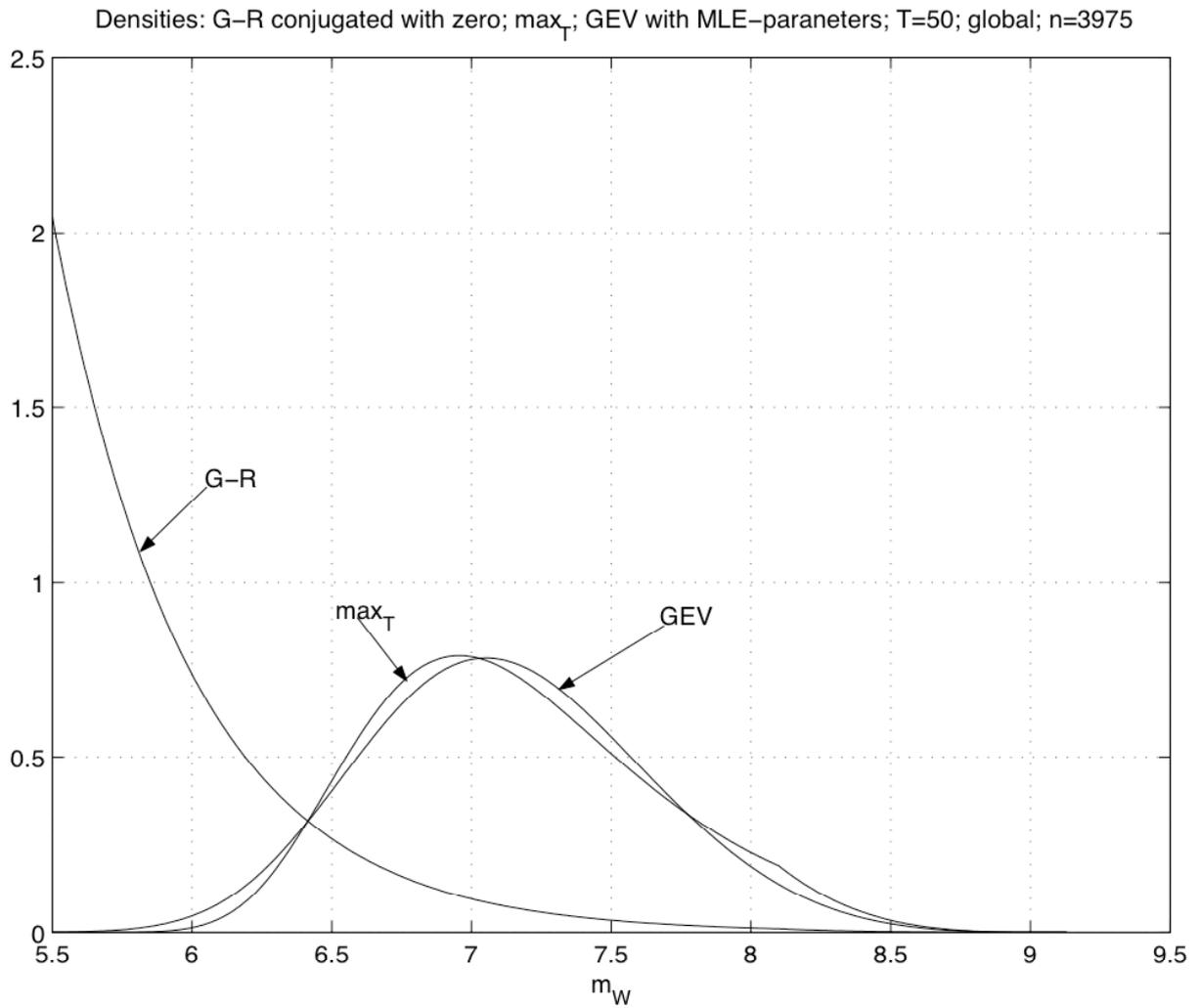

Fig.6: In linear scale, a two-branch density (G-R) is shown, then the resulting density of *T*-maxima calculated from equation (7) referred to as `max_T' in the figure and the approximating GEV density. In this example, the parameters of the two-branch generating model defined in equations (C1-C3) of Appendix C are *b=2.12, m_l=8.1*. The parameters of the approximating GEV distribution are: *m = 6.94, s = 0.482, ξ = -0.219, M_max = 9.13*.

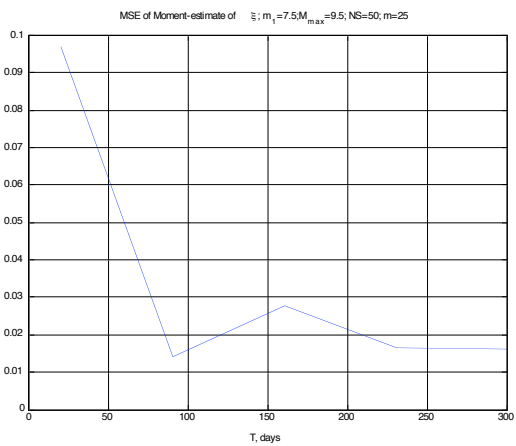

Fig. 7a

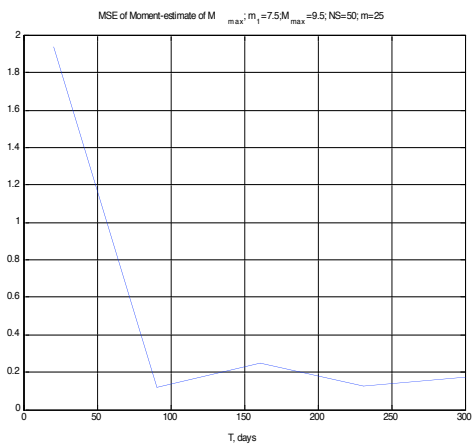

Fig. 7b

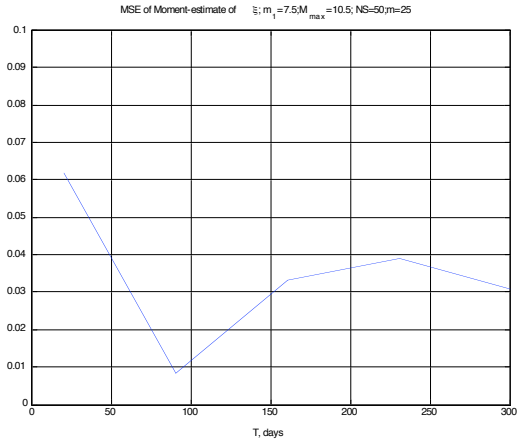

Fig. 7c

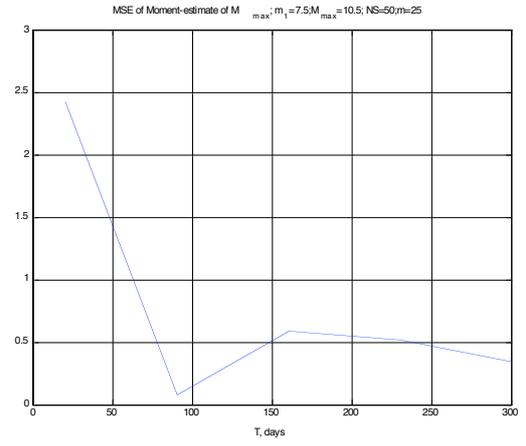

Fig. 7d

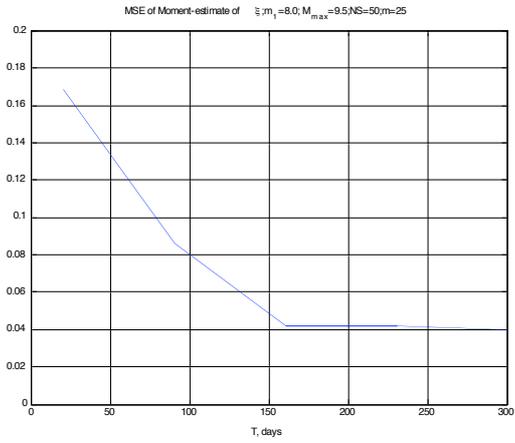

Fig. 7e

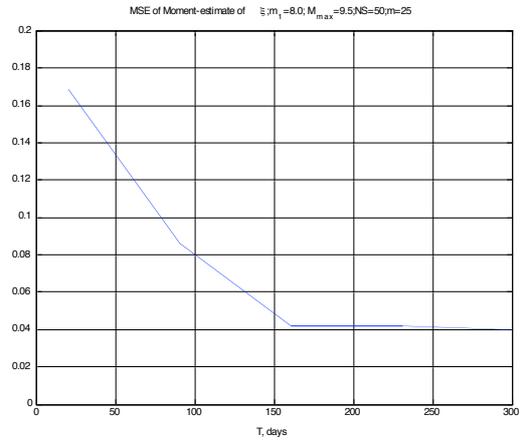

Fig. 7f

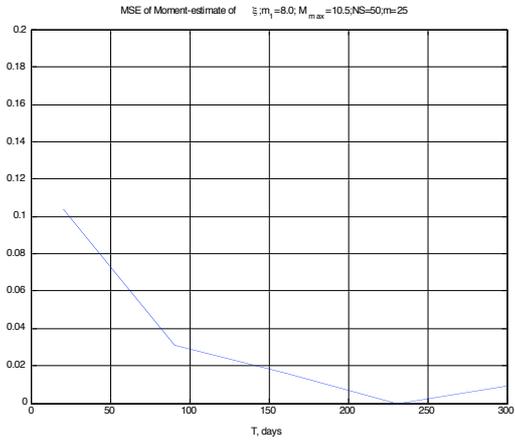

Fig.7g

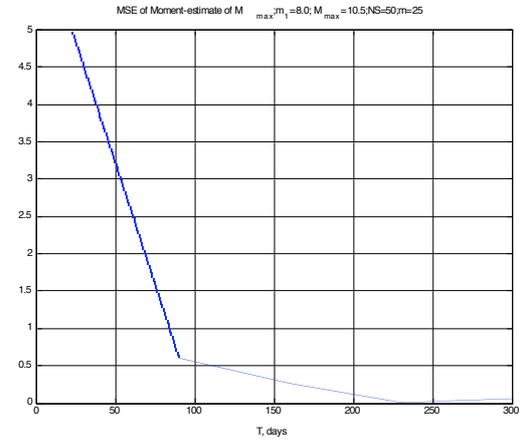

Fig. 7h

Fig. 7. Mean Square Errors of the Moment estimates of the parameter $\xi$ *(7a, 7c, 7e, 7g)* and the parameter $M_{max}$ *(7b, 7d, 7f, 7h)* obtained with the reshuffling procedure (number of reshuffling *NS = 50*; number of simulations $n_s$ *= 25*) on the Harvard catalog of shallow events (*01.01.77-18.12.04; h ≤ 70 km; $M_w$ ≥  5.5; N = 3975 main shocks*). The parameters *(m₁, M_{max})* of the two-branch distribution are indicated in the headings.

### 3.2 Moment estimates of the GEV-parameters for the empirical Harvard catalogs and their corrections for scatter and systematic biases by the simulation method.

Applying the Moment method with $T=182.5$ and with $NS = 1000$ reshufflings to the GEV distribution with three unknown parameters $(m, \sigma, \xi_g)$ ) for the distribution of $T$-maxima of the empirical global Harvard catalog ($01.01.77$-$18.12.04$; $m_W \geq 5.5$; $h \leq 70$ km; main shocks, $N=3975$), we get the following estimates:

$$\widetilde{m} = 7.49; \quad \widetilde{\sigma} = 0.381; \quad \widetilde{\widetilde{\xi}}_g = -0.320. \tag{17}$$

The parameter $M_{max}$ was estimated through the sample median of the set of estimates obtained with the reshuffled catalogs:

$$med(\widetilde{M}_{max}) = 8.70. \tag{18}$$

We estimated the scatter of these estimates within the set of reshuffled catalogs (recall that this is only a fraction of the full scatter since, for a unique real sample, one cannot estimate directly the global scatter) by calculating the corresponding std, denoted as $\sigma_m^{(r)}, \sigma_s^{(r)}, \sigma_{\xi}^{(r)}$. We obtained

$$\sigma_m^{(r)} \cong 0.031; \qquad \sigma_s^{(r)} \cong 0.025; \qquad \sigma_{\xi}^{(r)} \cong 0.051. \tag{19a}$$

For the parameter $M_{max}$, we took as a scatter measure the pair of sample quantiles at the probability levels 16% and 84% corresponding in the Gaussian case to the mean value ±std:

$$q_{Mmax}(0.16) = 8.57; \quad q_{Mmax}(0.84) = 8.83. \tag{19b}$$

For the sample 95% confidence interval, corresponding to quantiles 2.5% and 97.5%, we get:

$$q_{Mmax}(0.025) = 8.45; \quad q_{Mmax}(0.975) = 9.02.$$

We stress that (19a),(19b) characterize only a fraction of the full scatter. The histogram of $M_{max}$ for 1000 reshuffling is shown on Fig.8. One can see that just the contribution due to the reshuffling procedure already gives rise to a significant variation of the estimates of this parameter, which can take values going up to 9.25.

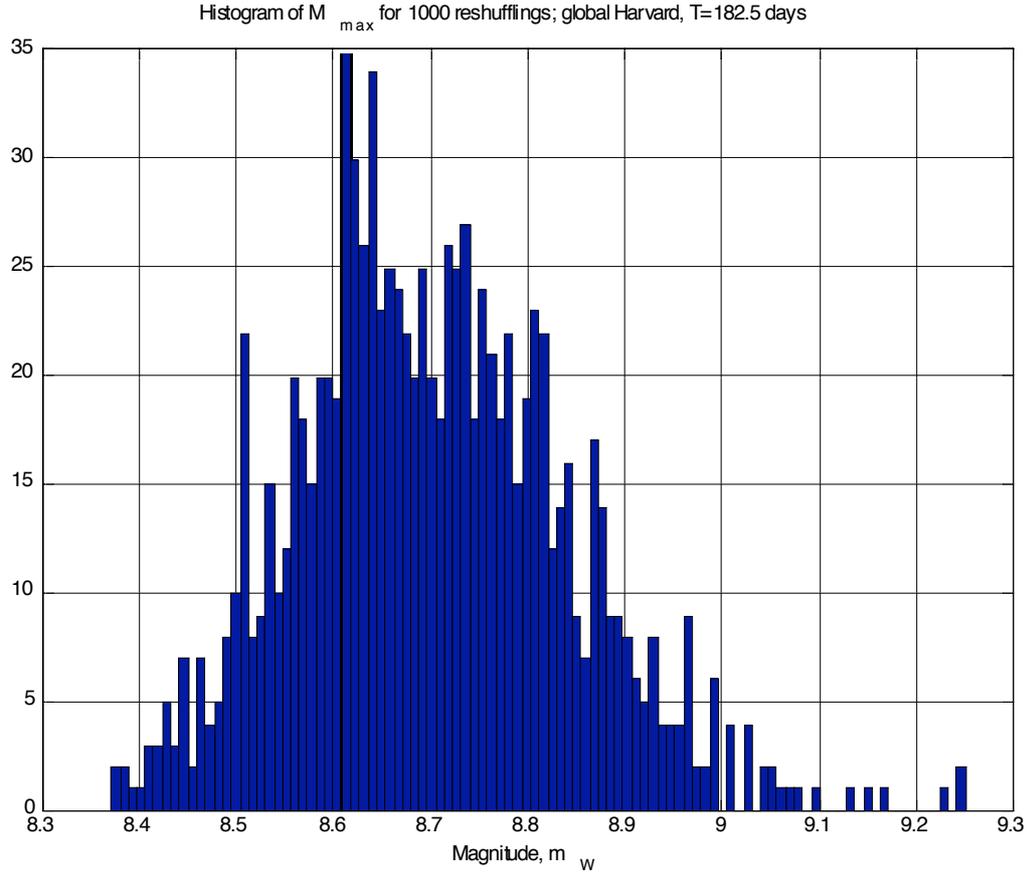

Fig. 8. Histogram of $M_{max}$ for *1000* reshuffled catalogs; $T = 182.5$ days; Harvard catalog of shallow events (*01.01.77-18.12.04; $h \leq 70$ km; $M_w \geq 5.5$; $N = 3975$ main shocks*).

We know from the simulation experiments described above and in the Appendix C that these estimates have a statistical scatter and some systematic bias depending both on $T$ and on the unknown true values of these parameters. We are going to evaluate these characteristics by the simulation method. For this purpose, we simulate the full procedure of the estimation starting from an artificial catalog generated by the two-branch model (C1-C3) with parameters *($m_1$, $M_{max}$)*, the parameter $b$ is fixed at the value *b=2.1*, all other parameters being determined through *($m_1$,$M_{max}$)*. As a representative set of parameter pairs, we consider again the four pairs (15). We fixed $T = 182.5$ days. The resulting statistical characteristics of the Moment estimates are collected in Table 2.

Table 2. Statistical characteristics of the Moment estimates of artificial samples generated by the two-branch model with parameters *($m_1$, $M_{max}$)* corresponding to the four pairs (15); *T=182.5* days; number of reshufflings *NS = 100*; number of simulations $n_s = 50$.

| $m_1$= 7.5; $M_{max}$ = 9.5 | |
|---|---|
| $\sigma_m =std(\widetilde{m}) = 0.0423$ | |
| $\sigma_s = std(\widetilde{\sigma}) = 0.0270$ | |
| $\sigma_{\widetilde{\xi}} =std(\widetilde{\widetilde{\xi}}) = 0.0618$ | $bias(\widetilde{\widetilde{\xi}}) = 0.0086$ |
| $median(\widetilde{M}_{max})$=9.62 | $q_{Mmax}(0.16) = 8.99$;   $q_{Mmax}(0.84) = 10.37$ |
| $m_1$= 7.5; $M_{max}$ = 10.5 | |
| $\sigma_{m=} std(\widetilde{m}) = 0.0534$ | |
| $\sigma_s = std(\tilde{\sigma}) = 0.0327$ | |
| $\sigma_{\tilde{\xi}} = std(\tilde{\xi}) = 0.0586$ | $bias(\tilde{\tilde{\xi}}) = -0.0061$ |
| $median(\tilde{M}_{max}) = 10.51$ | $q_{Mmax}(0.16) = 9.44;\quad q_{Mmax}(0.84) = 12.94$ |
| | |
| **$m_1 = 8.0;\ M_{max} = 9.5$** | |
| $\sigma_m = std(\tilde{m}) = 0.0484$ | |
| $\sigma_s = std(\tilde{\sigma}) = 0.0322$ | |
| $\sigma_{\tilde{\xi}} = std(\tilde{\xi}) = 0.0611$ | $bias(\tilde{\tilde{\xi}}) = 0.0842$ |
| $median(\tilde{M}_{max}) = 10.23$ | $q_{Mmax}(0.16) = 9.51;\quad q_{Mmax}(0.84) = 12.65$ |
| | |
| **$m_1 = 7.5;\ M_{max} = 10.5$** | |
| $\sigma_m = std(\tilde{m}) = 0.0512$ | |
| $\sigma_s = std(\tilde{\sigma}) = 0.0325$ | |
| $\sigma_{\tilde{\xi}} = std(\tilde{\xi}) = 0.0574$ | $bias(\tilde{\tilde{\xi}}) = 0.0322$ |
| $median(\tilde{M}_{max}) = 11.26$ | $q_{Mmax}(0.16) = 9.84;\quad q_{Mmax}(0.84) = 13.41$ |

We observe that the std of all three parameters *(m,s,ξ)* can be taken approximately constant around the values

$$\sigma_m \cong 0.05; \qquad \sigma_s \cong 0.03; \qquad \sigma_{\tilde{\xi}} \cong 0.06. \qquad (20)$$

We infer that the std of the Moment estimates (17) in the real sample are close to (20). As to the bias of the ξ-estimate, we see that it is negligible (as compared to its corresponding std) for the first two sets of *(m₁, Mmax)*, and comparable to the corresponding std for the third and fourth sets. Since we have no information about the possible true values of *(m₁, Mmax)*, we prefer to introduce no correction for the bias of the ξ-estimate, keeping in the mind the possibility for the existence of a bias which can be as large as 0.085.

Let us now discuss the scatter of the moment-estimate of the parameter *$M_{max}$*. As it was remarked above, this estimate is rather unstable as compared with the estimates of the other parameters. Therefore, it is more appropriate to estimate its scatter by the order sample statistics, namely, the sample quantiles, which are more robust statistically. We thus took the sample "spread" corresponding to one std in the Gaussian case:

*Upper Spread (US) = q(0.84) - median .*

We averaged these spreads over the four sets of pairs *(m₁, Mmax)* used in Table 2, and obtained the following average spread (the robust analog of the std) for the parameter *$M_{max}$*:

$$US = 1.93 \cong 2.0; \qquad (21)$$

Then, using the sample median (18) for the real Harvard sample and the spread (21), we can provide the following upper confidence bound $\hat{q}(0.84)$ at the probability level *84%* (which is the analog of one-std upper bound for the Gaussian distribution):

$$\hat{q}(0.84) = 8.70 + 2.0 = 10.7 .$$

This value is considerably larger than the fractional upper bound $q_{Mmax}(0.84) = 8.83$ given by (19b).

Our final estimates of the GEV-parameters with their scatter characteristics for the global Harvard catalog are the following:

$$\widetilde{m} = 7.5 \pm 0.05; \qquad ; \ \widetilde{\sigma} = 0.38 \pm 0.03; \qquad \widetilde{\widetilde{\xi}}_g = -0.32 \pm 0.06 ; \qquad (22)$$

$$med(\widetilde{M}_{max}) = 8.7; \quad \widetilde{q}_{M\,max}(0.84) = 10.7. \qquad (23)$$

Let us now turn to the determination of the estimators of the two better and more stable characteristics introduced in section 2.3, the quantile $Q_T(q)$ (defined in equation (12)) and the probability $\rho_T(m^*)$ (defined in equation (13)) for the GEV distribution fitted to the empirical distribution of the $T$-maxima. For the global Harvard catalogue ($01.01.77$-$18.12.04$; $m_W \geq 5.5$; $h \leq 70$ km; main shocks, $N=3975$), we get (for the optimal value $T=182.5$ and with $NS=1000$ reshufflings) the following Moment-estimates:

$$Q_T(0.98) = 8.34; \qquad log_{10}(\rho_T(8)) = -0.799; \qquad (24)$$

In order to evaluate the statistical scatter of these estimates, we use again the two-branch model with the four sets of parameters (15). We obtain the following sample estimates of the std with $NS=100$ reshufflings and $n_s=100$ simulations:

$m_1, = 7.5;\ M_{max} = 9.5;$     $std(Q_T(0.98)) = 0.121;\ std(log_{10}(\rho_T(8))) = 0.114;$
$m_1, = 7.5;\ M_{max} = 10.5;$    $std(Q_T(0.98)) = 0.180;\ std(log_{10}(\rho_T(8))) = 0.097;$
$m_1, = 8.0;\ M_{max} = 9.5;$     $std(Q_T(0.98)) = 0.145;\ std(log_{10}(\rho_T(8))) = 0.081;$
$m_1, = 8.0;\ M_{max} = 10.5;$    $std(Q_T(0.98)) = 0.172;\ std(log_{10}(\rho_T(8))) = 0.080;$

The relative scatter of these characteristics is rather small, so that we can use their average values for the evaluation of the scatter of the real sample estimates (24):

$$std(Q_T(0.98)) \cong 0.16; \qquad std(lg_{10}(\rho_T(8))) \cong 0.1. \qquad (25)$$

For the global Harvard catalog, we thus obtain the following estimates:

$$Q_T(0.98) = 8.34 \pm 0.16 ; \qquad log_{10}(\rho_T(8)) = -0.80 \pm 0.1. \qquad (26)$$

It is important to notice that the scatters of $Q_T(0.98)$ and $log_{10}(\rho_T(8))$ are much smaller than the scatter of the $M_{max}$-estimate, whose upper 84%-bound differs from the median by 2 (see equation (23)), compared with 0.2 for $Q_T(0.98)$. ), Fig. 9a-c illustrate the stability of the former two estimates with respect to the latter by showing three histograms of the corresponding sample estimates calculated for the two-branch model with parameters given by the third set in (15): $m_1, = 8.0;\ M_{max} = 9.5;\ \xi = -0.2410;\ s=0.3614$. Observe that the $M_{max}$-estimate goes up to the value 20!

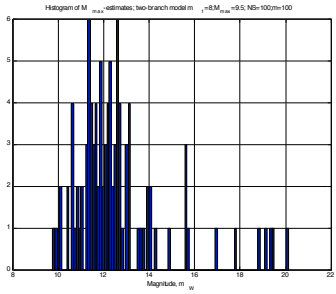 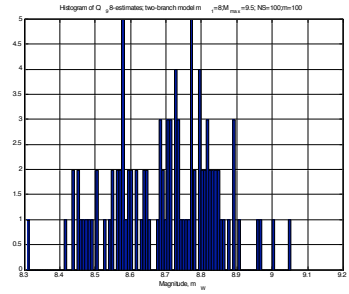 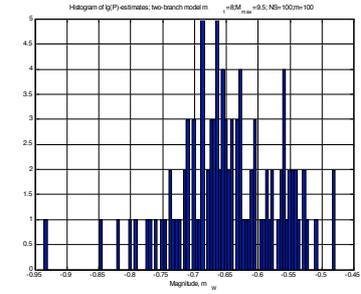

| Fig. 9a | Fig. 9b | Fig. 9c |

Fig.9 Histograms of the estimates of $M_{max}$ (9a), $Q_{0.98}$ (9b), and $log_{10}(P\{ max_T >8.0\})$ for the two-branch model with parameters $m_1, = 8.0$; $M_{max} = 9.5$. NS = 100 reshufflings; $n_s = 100$ simulations.

### 3.3 Changes in the estimates caused by taking into account the great Sumatra earthquake 24.12.04 , $m_W = 9.0$.

The occurrence of the great Sumatra earthquake offers an excellent opportunity for a so-to-speak out-of-sample test of our results obtained until now. In this section, we demonstrate the impact of the great Sumatra EQ 24.12.04 on the stability of our estimation procedure. Specifically, we take for the analysis the time period 01.01.1977- 16.06.2006, with a total duration of 10759 days. We have applied to this period the procedure of declustering described in Appendix A1 and obtained 4193 main shocks. We have used the same value of $T$ as earlier, namely $T = 182.5$ days. We obtained the following ME estimates of the GEV-parameters fitted to this new sample:

$$\hat{m} = 7.49; \quad \hat{\sigma} = 0.381; \quad \hat{\xi} = -0.178. \qquad (27)$$

As compared to our earlier estimates (17), the estimates of the parameters $m, \sigma$ practically did not change whereas the $\xi$-estimate has changed dramatically. The same can be said for the estimates of the parameter $M_{max}$ for which we obtain the following *16% - 50% - 84%* sample percentiles:

$$q_{Mmax}(0.16) = 9.27; \quad q_{Mmax}(0.50) = 9.67; \quad q_{Mmax}(0.84) = 10.31. \qquad (28)$$

Recall that this scatter is due to the reshuffling procedure (*NS = 1000* reshufflings) which provides only a lower bound of the full scatter which must includes the variation from sample to sample. We see that the new median estimate $q_{Mmax}(0.50) = 9.67$ exceeds the previous one $q_{Mmax}(0.50) = 8.70$ obtained for the period *01.01.77-18.12.04* by almost one unit. This illustrates once more the instability of the estimation of the parameter $M_{max}$. For comparison with Fig.8, Fig.10 shows the new histogram of the reshuffling estimates of $M_{max}$. Compared with Fig. 8, the histogram of $M_{max}$ has changed dramatically. For instance, the previous upper bound was *9.25* whereas the new one is *15.4*! This figure confirms once more the instability of the estimation of $M_{max}$.

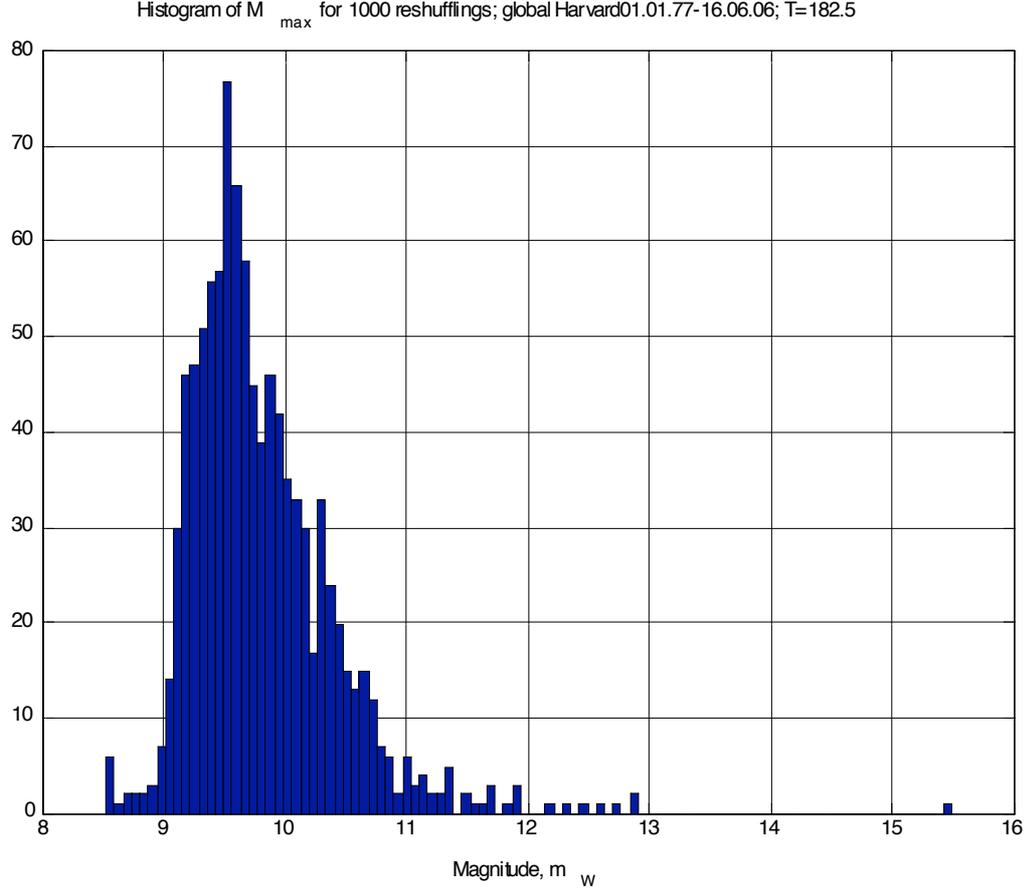

Fig. 10: Same as Fig. 8 but for the time period 01.01.1977- 16.06.2006, $h \leq 70 \ km; \ M_w \geq 5.5$, with a total duration of 10759 days and 4193 main shocks, which includes the great Sumatra earthquake *24.12.04 , $m_W = 9.0$.*

Let us now turn to the parameters introduced in section 2.3 characterizing the tail behavior of the magnitude distribution, which we have proposed and argued for on the basis of their good and stable behavior. They consist in the quantile $Q_T(q)$ (defined in equation (12)) and the probability $\rho_T(m*)$ (defined in equation (13)) for the GEV distribution fitted to the empirical distribution of the $T$-maxima. We obtain the following values for $Q(0.98)$ and $log_{10}(\rho_T(8.0) \ )$ for the new sample:

$$Q^{new}(0.98) = 8.57; \quad log_{10} \ ( \ \rho_T^{new}(8.0) \ \ ) = - \ 0.705 \ , \quad (29)$$

which should be compared with old estimates (26): *Q(0.98) = 8.34 ± 0.16; $log_{10}(\rho_T(8.0) \ )$ = -0.80 ± 0.1.* The new estimates differ only by approximately one *std* from the previous ones. This can be considered as quite admissible, taking into account the expected impact of the outstanding event of the great Sumatra earthquake.

## 4   Summary and conclusions

We have developed a new method for the statistical estimation of the tail of the distribution of earthquake sizes. The method focused on $max_T$ – maximum magnitude observed in successive intervals of length $T$. We have suggested a new parametric model for the distribution of main shock magnitudes, which we referred to as the two-branch model (C1-C3) in Appendix C.  We found that the number of main events in the catalog (*N = 3975*) turned out to be insufficient in order to use a direct estimation of the parameters of this model, due to the inherent instability of the estimation problem. This instability can be explained by the fact that only a small fraction of the empirical data populates the second branch described by three unknown parameters ($m_1$ ; $\alpha$ ; $\sigma$ ). In this vein, *Pisarenko and Sornette (2004)* estimated by an independent method that, for the global Harvard catalogs for shallow earthquakes, about 17 events in the tail of the distribution departs significantly from the pure G-R distribution. Thus, even a sample size of about $N = 4000$ can be insufficient for the reliable estimation of these parameters and a degenerate behavior of the model as explained in Appendix C can occur. This instability is made even more apparent, if need be, by our comparison of the estimates of $M_{max}$ over the period *01.01.77-18.12.04* with that over the period *01.01.1977- 16.06.2006* which includes the great Sumatra EQ 24.12.04, $m_W = 9.0$: we find that the median of the GEV distribution of $M_{max}$ increases almost by one unit magnitude when including the impact of the Sumatra earthquake, while the interval of maximal admissible $M_{max}$ jumps from *9.25* to *15.4*!

This led us to develop an indirect method for the estimation of the parameters describing the tail of the distribution of earthquake magnitudes, namely, through $max_T$ that we called the set of *T*-maxima, i.e., the *n* maximum magnitudes in *n* windows of duration equal to *T* days. We have confirmed the known fact that, for samples of small and intermediate size *( n ≤ 50)*, the Moment estimates of the GEV-parameters are more efficient than the Maximum Likelihood estimates (see *Hosking et al. 1985, Coles and Dixon 1999*). We have established that in our estimation problem for small sample size the Moment estimates are slightly more efficient than the Probability-Weighted Moment estimates (see *Christopeit 1994, Embrechts et al. 1997)*. We have suggested a method for the determination of the optimal choice of the *T*-value minimizing the Mean Square Error of the estimation of the form parameter of the GEV distribution approximating the sample distribution of *T*-maxima. This approach makes possible the Moment-estimation of the form parameter of the tail distribution, although a considerable uncertainty of the estimation procedure still persists. Perhaps, more representative catalogs of earthquake magnitudes can help increase the sample size and make the estimation more reliable. But generalizing the suggested methods to magnitude catalogs demands some further work, in particular to take into account the discrete character of the magnitudes and the distortion associated with the time evolution of the completeness of the magnitude catalogs. Perhaps, even some historical catalogs based on the registration of seismic intensity can be used to further increase the sample size, but this question needs a more detailed study.

We now summarize our main results. We have derived the following Moment-estimates for the GEV distribution of earthquake magnitudes fitted to the sample distribution of *T*-maxima of the global Harvard catalog of main shocks over the time period 1977-2004 and 1977-2006:

$$\widetilde{m} = 7.49 \pm 0.049; \qquad ; \ \widetilde{\sigma} = 0.381 \pm 0.031 \ \text{(both periods)} \tag{30a}$$

$$\widetilde{\widetilde{\xi}}_g = -0.320 \pm 0.060 \ (01.01.77\text{-}18.12.04) \quad ; \ \hat{\widetilde{\xi}} = -0.178 \ (01.01.77\text{-} 16.06.06) \tag{30b}$$

$$med(\widetilde{M}_{\max}) = 8.70 \ (01.01.77\text{-}18.12.04); \ \ med(\widetilde{M}_{\max}) = 9.67 \ (01.01.77\text{-} 16.06.06) \tag{30c}$$

$$\widetilde{q}_{M\max}(0.84) = 10.70 \ (01.01.77\text{-}18.12.04); \ \widetilde{q}_{M\max}(0.84) = 10.31 \ (01.01.77\text{-} 16.06.06 \tag{30d}$$

We have suggested two stable statistical characteristics of the tail of the distribution of earthquake magnitudes: the quantile $Q_T(q)$ of a high probability level *q* for the *T*-maxima, and the probability of exceedence of a high threshold magnitude $\rho_T(m^*) = P\{\ m_k \geq m^*\}$. The statistical estimates of these characteristics are more stable than the estimate of $M_{max}$ . We obtained the following sample estimates for the global Harvard catalog:

$$Q_T(0.98) = 8.34 \pm 0.16 \ ; \qquad log_{10}(\rho_T(8)) = -0.80 \pm 0.1; \quad (\rho_T(8)) = 0.13\text{-}0.2) \tag{31}$$

The relative scatter of these estimates is much smaller than the scatter of the $M_{max}$-estimate and remains robust even when included the impact of the great Sumatra earthquake. We would like to note that using the set of $m_{max}(T)$ as suggested in our paper with the estimation of the GEV-parameters permits to evaluate all statistical characteristics (including $Q(q)$ and $\rho(m^*)$ ) related to the future time intervals that are multiples of $T$, since for $T_1 = r \cdot T$ the distribution function of $m_{max}(T)$ is $[GEV(x \mid m,s,\xi)]^r$.

**Acknowledgements.**

The work was supported partially (V.F.Pisarenko, M.V.Rodkin) by the Russian Foundation for Basic Research, grant 05-05-65276.

**APPENDIX A: Elimination of aftershocks.**

In order to get more reliable estimates, it is necessary decluster catalogs to remove the strongest time dependence between so-called aftershocks and main shocks so as to approach better the condition of independence necessary for the applications of the theorems of extreme value theory.

The method of declustering that we use is standard. We start by identifying the largest event in a catalog, whose time, location and magnitude are denoted as (time $t$, location $(\phi_0, \lambda_0)$, magnitude $M$). Then, we exclude all events in the time-space window:

$$(t; \ t +10^{-0.31+0.46\,M});$$
$$R(\phi, \lambda; \phi_0, \lambda_0) \leq 10^{-0.85+0.46\,M}; \qquad\qquad (A1)$$

where $R$ is the distance in km between points $(\phi, \lambda)$ and $(\phi_0, \lambda_0)$. The window (A1) was taken from (*Knopoff et al.1982*). After the first elimination, we identify the next largest event of the remaining earthquakes (excluding the previous one already accounted for). And we apply the same pruning with the same rule for the space-time window associated with this second largest event of remaining earthquakes. And we iterate until the algorithm stops.

If this algorithm for the elimination of aftershocks was perfect, we should obtain a Poisson distribution of main shocks (provided of course that the so-called main shocks can be assumed to be independently occurring according to a constant Poisson rate). Therefore, a natural procedure to check for the quality of the algorithm eliminating the aftershocks is to test whether the declustered catalog obeys the Poisson statistics. For this, we use two characteristics:

1. The standard Kolmogorov distance between the sample distribution of occurrence times of events (where the times are normalized to unity);
2. The ratio $\mu$ = (sample variance)/(sample mean) for the numbers of events in intervals of length $T$ (We took $T = 50$ days for the experiments presented below). Under the null hypothesis of a Poisson distribution, this ratio is distributed approximately as a Chi-square random variable with $K-1$ degrees of freedom, where $K$ is the number of such intervals of length $T$ *(Cox and Lewis, 1966)*. In our experiments, $K=203$.

We used the Harvard catalog of seismic moments over the time period 01.01.1977-16.05.2006. Let us consider first the magnitude range $m_W \geq 5.0$. The sample distribution of normalized occurrence times (which for a Poisson process should be uniform) is shown on fig. A1. One can see clear deviation from the Poisson process. The inset also shows that there is a significant trend in the number of recorded earthquakes over time, probably resulting from the improvement in the recording system. In addition, the large peak at the end of the time series of event number, followed by a sharp drop, is associated with the Sumatra earthquake with $m_W =9.0$ of December 24, 2004. The two statistics obtained for this time series are $\mu = 13.97$ (probability of exceeding this value under the Poisson hypothesis is practically zero; for a Poisson process, theoretically $\mu = 1$). We also obtain $KD=13.97$ (outside of tabulated results, i.e., the probability of exceeding this value under the Poisson hypothesis is also practically zero). We thus conclude that the null Poisson hypothesis is totally rejected for this catalog over the time period 01.01.1977-16.05.2006 and magnitude range $m_W \geq 5.0$.

For the same magnitude range, when we restrict to the time interval 01.01.77-20.12.04 to remove the influence of the Sumatra EQ with $m_W =9.0$ which occurred on 24.12.04, we obtain $\mu = 7.90$, $KD=14.62$ (outside of table) again very strongly rejecting the null hypothesis of a Poisson process.

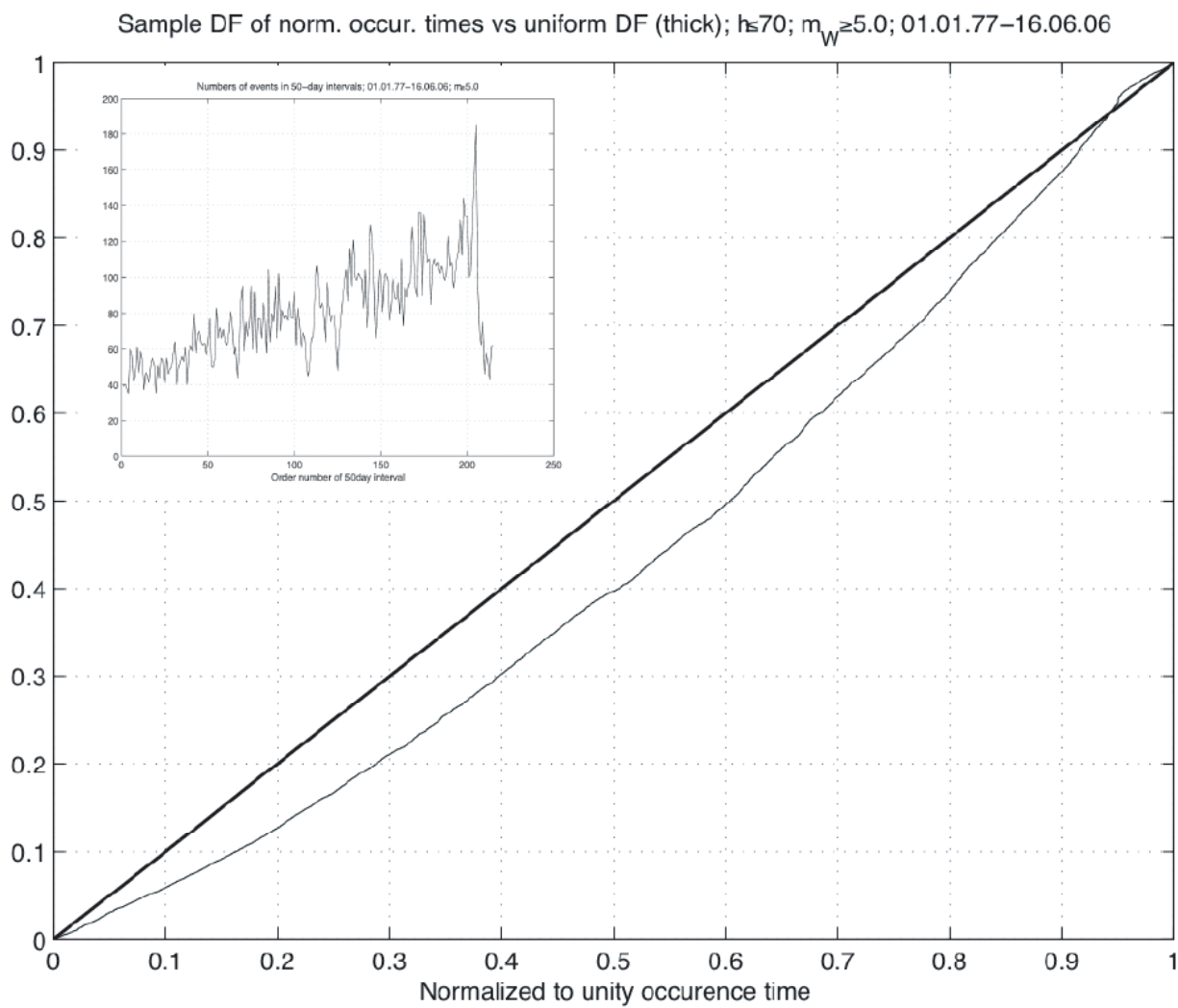

Fig.A1: Sample distribution of the normalized occurrence times of earthquakes of magnitude $m_W \geq 5.0$ in the Harvard catalog of seismic moments over the time period 01.01.1977-16.05.2006. For a Poisson distribution, the data should align on the diagonal. The inset shows the numbers of events in sequential 50-day intervals.

We now consider the Harvard catalog of seismic moments over the time period 01.01.77-20.12.04 and for the magnitude range $m_W = 5.5$. Fig. A2 exhibits a significant improvement over Fig. A1: the trend in the inset almost disappears but there are still significant deviations from the Poisson distribution, with obvious time clustering associated with aftershocks. This translates into the following statistics $\mu = 2.14$ (probability of exceeding this value under Poisson hypothesis is practically zero), $KD=3.15$ (still outside of table), thus rejecting again strongly the Poisson null hypothesis.

Fig. A3 shows the results obtained using the algorithm removing the aftershocks described above. From $n=7521$ shallow events in the time interval 01.01.77-20.12.04 with $m_W \geq 5.5$, we retain *3975* main events *(3975/7521 $\cong$ 52.9%* so that approximately half of the events are triggered according to this algorithm), and we obtain $\mu = 1.12$ (probability of exceeding this value under the Poisson hypothesis is *10.7%), KD=0.836* (probability of exceeding this value is about 50%). The inset exhibits no visible trend. These different diagnostic suggest that the algorithm removing the aftershocks can be considered as satisfactory.

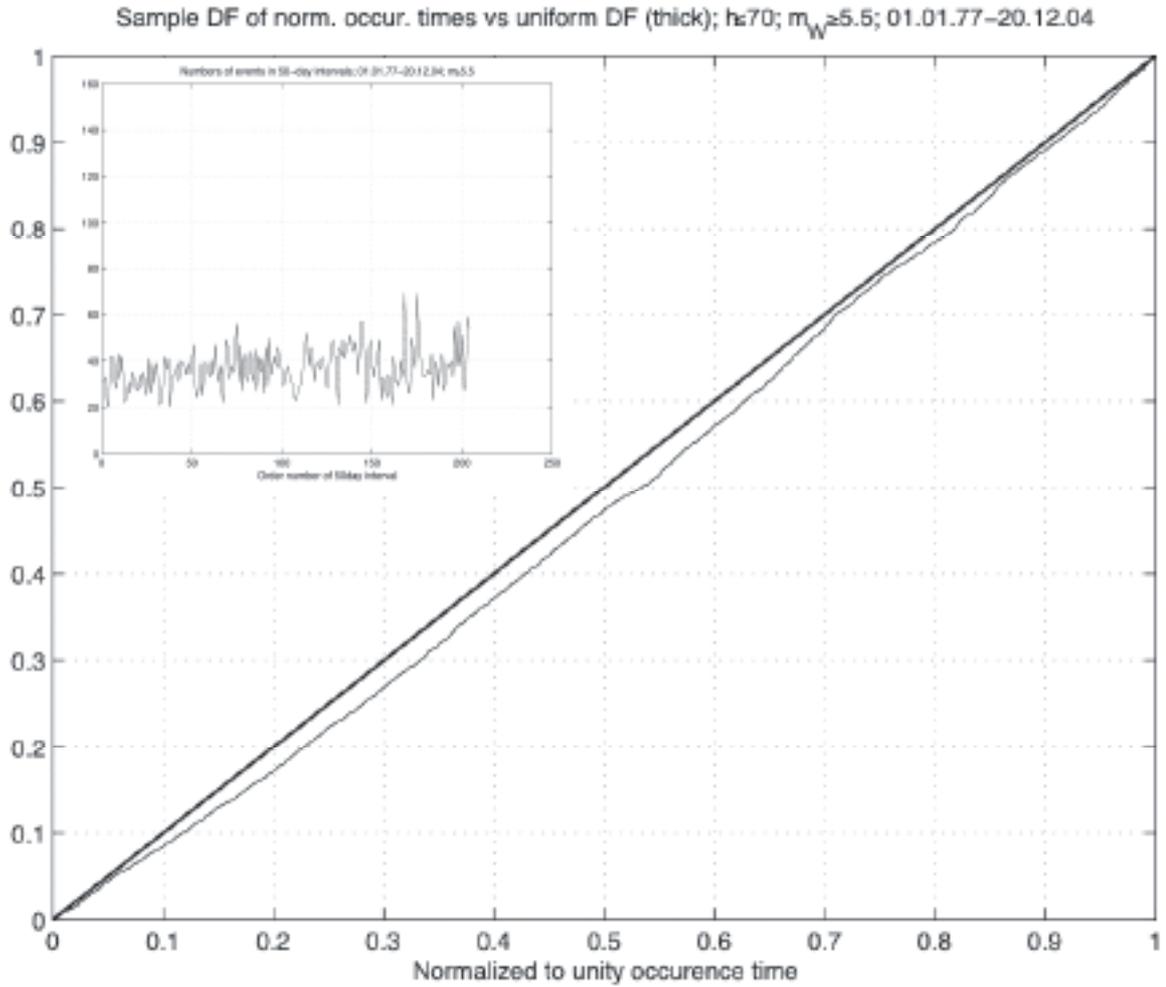

Fig.A2: Sample distribution of the normalized occurrence times of earthquakes of magnitude $m_W \geq 5.5$ in the Harvard catalog of seismic moments over the time period 01.01.77-20.12.04. For a Poisson distribution, the data should align on the diagonal. The inset shows the numbers of events in sequential 50-day intervals.

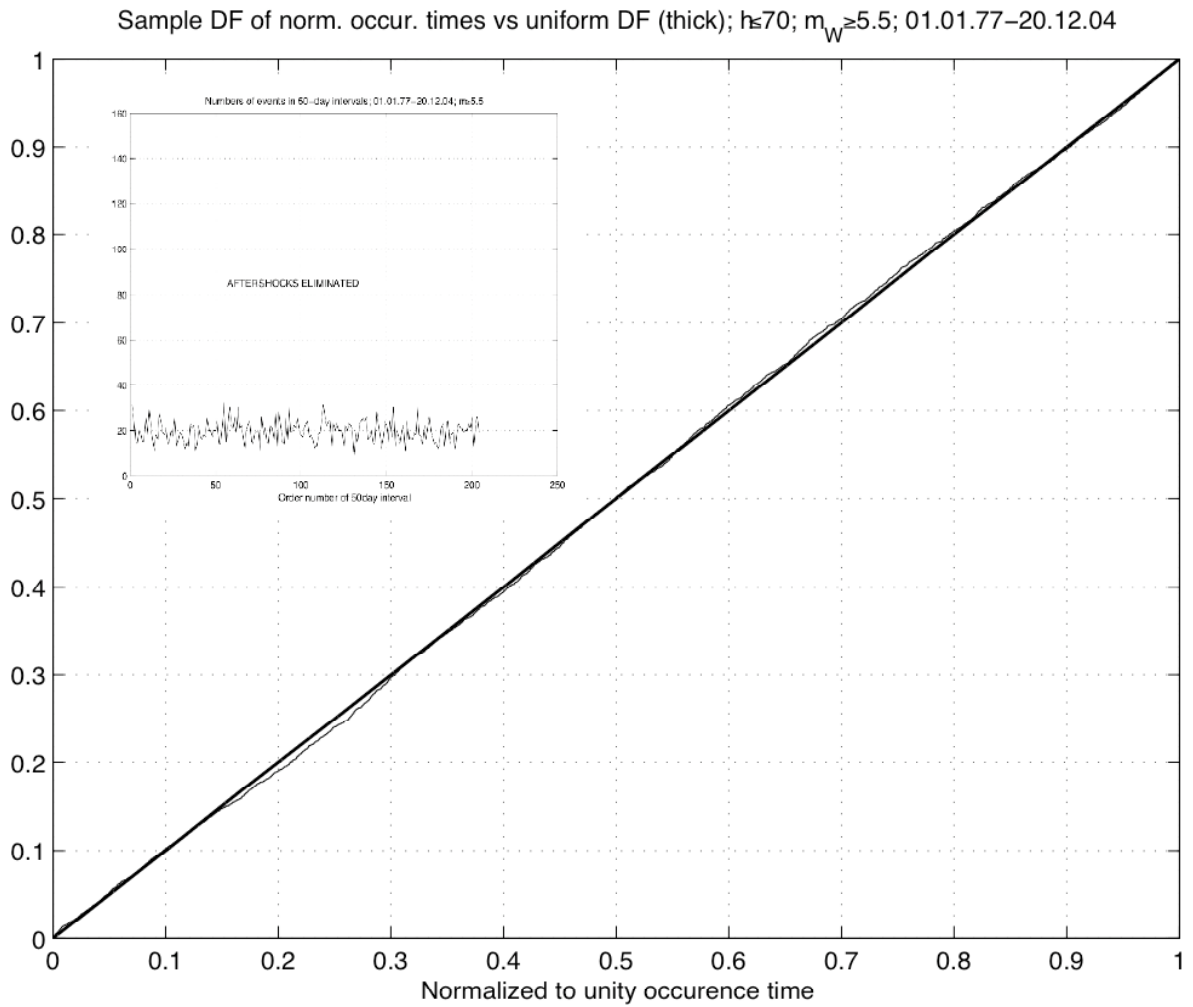

Fig.A3: Sample distribution of the normalized occurrence times of declustered "main shock" earthquakes of magnitude $m_W \geq 5.5$ in the Harvard catalog of seismic moments over the time period 01.01.77-20.12.04 obtained by using the declustering algorithm described in the text. For a Poisson distribution, the data should align on the diagonal. The inset shows the numbers of "main shock" events in sequential 50-day intervals.

**APPENDIX B: Comparison of three methods of estimation of the parameters (*m, s, ξ*) of the GEV distribution function: the Maximum Likelihood Estimates (*MLE*), the Moment Estimates (*ME*) and estimates based on Probability-Weighted Moments.**

Consider a sample *(x₁,...,xₙ)* generated with the distribution function *GEV(x |m,s,ξ)*:

$$GEV(x \mid m,s,\xi) = exp\left(-[1 + \frac{\xi(x-m)}{s}]^{-1/\xi}\right); \qquad 1 + \frac{\xi(x-m)}{s} > 0; \quad \xi < 0. \qquad (B1)$$

We consider the problem of the estimation of the parameters *(m,s,ξ)*, with three well-known methods of statistical estimation of parameters: the Maximum Likelihood Estimates (*MLE*), the Moment Estimates (*ME*) and estimates based on Probability-Weighted Moments (*PWM*, see e.g. *Embrechts et al. 1997*). Since the support of the distribution *GEV(x)* depends on the unknown parameters as defined in (B1), the regularity of the *MLE* and its asymptotic efficiency properties are not obvious. In (*Smith 1985*) it was shown that the *MLE* are asymptotically regular at least for *ξ > -0.5*. But for small and moderate sample sizes *(n ≅ 10 ÷ 100)*, the *MLE* turned out to be less efficient than the *ME* and the *PWM* estimate (see *Christopeit 1994, Coles and Dixon 1999*). We have compared these three *MLE, ME* and *PWM* estimates on small and moderate samples and confirm that the *MLE* is less efficient than both other competing methods, and besides, the *ME* is slightly superior than the *PWM* estimate.

Le us recall the definitions of the *ME* and *PWM* estimate. Let us denote the first three centered sample moments as follows:

$$M_1 = \frac{1}{n}\sum_{k=1}^{n} x_k; \qquad M_2 = \frac{1}{n}\sum_{k=1}^{n} (x_k - M_1)^2; \qquad M_3 = \frac{1}{n}\sum_{k=1}^{n} (x_k - M_1)^3. \qquad (B2)$$

The corresponding theoretical moments for the *GEV*-distribution (B1) are:

$$\mu_1 = m - s/\xi + \frac{s}{\xi} Gamma(1 - \xi); \qquad (B3)$$

$$\mu_2 = s^2/\xi^2 \left[ Gamma(1 - 2\xi) - (Gamma(1 - \xi))^2 \right];$$

$$\mu_3 = s^3 \left[-2(Gamma(-\xi))^3 - \frac{6}{\xi} Gamma(-\xi) Gamma(-2\xi) - \frac{3}{\xi^2} Gamma(-3\xi) \right];$$

In the *ME* method, we identify the theoretical moments with their sample analogs and solve the resulting equations with respect to the unknown parameters:

$$m - \frac{s}{\xi} + \frac{s}{\xi} Gamma(1 - \xi) = M_1 ; \qquad (B4)$$

$$\frac{s^2}{\xi^2} \left[ Gamma(1 - 2\xi) - (Gamma(1 - \xi))^2 \right] = M_2 ; \qquad (B5)$$

$$s^3 \left[-2(Gamma(-\xi))^3 - \frac{6}{\xi} Gamma(-\xi) Gamma(-2\xi) - \frac{3}{\xi^2} Gamma(-3\xi) \right] = M_3 . \qquad (B6)$$

We can solve *explicitly* equations (B4)-(B5) with respect to *m,s*:

$$s_{ME} = [ \frac{M_2 \xi^2}{Gamma(1-2\xi) - (Gamma(1-\xi))^2} ]^{1/2} \qquad (B7)$$

$$m_{ME} = \frac{s}{\xi} - \frac{s}{\xi} Gamma(1-\xi) + M_1 . \qquad (B8)$$

Inserting (B7) into equation (B6), we have only one 1D equation for the numerical search of the solution for $\xi$. This can be effectuated by the most reliable and simple grid calculation.

In the *PWM* method, we identify the theoretical expectations

$$\gamma_k = E[ X \{GEV(X)\}^k ]; \qquad k=0,1,2; \qquad (B9)$$

with their sample analogs

$$g_k = \frac{1}{n} \sum_{j=1}^{n} x_j \{F_n(x_j)\}^k ; \qquad k=0,1,2, \qquad (B11)$$

where $F_n$ is the sample distribution function calculated from the sample $(x_1,...,x_n)$. The theoretical expectations (B9) are given by:

$$\gamma_k = \frac{m - (s/\xi)[1 - (k+1)^{-a} Gamma(1-\xi)]}{k+1} ; \qquad k=0,1,2. \qquad (B12)$$

Again, the solution of the system of equations, $\gamma_k = g_k$, $k=0,1,2$, can be reduced to the numerical search of the solution of the 1D equation:

$$\frac{3^{\xi} - 1}{2^{\xi} - 1} = \frac{3g_2 - g_0}{2g_1 - g_0}. \qquad (B13)$$

We denote this solution as $\xi_{PWM}$. Then, the two other *PWM*-estimates are expressed as follows:

$$s_{PWM} = \frac{(2g_1 - g_0)\xi_{PWM}}{(2^{\xi_{PWM}} - 1)Gamma(1-\xi_{PWM})} ; \qquad (B14)$$

$$m_{PWM} = g_0 + \frac{s_{PWM}(1 - Gamma(1-\xi_{PWM})}{\xi_{PWM}} . \qquad (B15)$$

The *MLE* are found as the parameter values $(m,s,\xi)$ maximizing the likelihood $L$:

$$L = \frac{1}{s^n} \prod_{j=1}^{n} [1 + \frac{\xi(x_j - m)}{s}]^{-1-1/\xi} exp\{ - [1 + \frac{\xi(x_j - m)}{s}]^{-1/\xi} \}. \qquad (B16)$$

Thus, the *MLE* involves a 3D numerical search, which is rather inconvenient from a computational point of view.

Let us now present the comparison between the results obtained with the three estimation methods, when applied to synthetic data sets of increasing sizes generated with parameters which are typical of what can be expected for real earthquakes. We generated samples of sizes *n = 10; 15; 25; 50; 200* with the parameters *m=7.5; s=0.4; $\xi$ = -0.2.* These parameters correspond, according to equation (8) to $M_{max} = m - s/\xi = 9.5$. We simulated 1000 samples for each size variant and determined numerically the statistical characteristics of each estimation: mean value and std (over 1000 realizations) for the parameters $(m, s, \xi)$, and the 16% and 84% sample quantiles for $M_{max}$. For the parameter $\xi$, we provide the Mean Square Error (*MSE*) that equals to $(bias^2 + std^2)^{1/2}$. The results are collected in Tables B1-B5.

Table B1. Comparison of the three methods of estimation for *n = 10*.

|  | *ME* | *MLE* | *PWM* |
|---|---|---|---|
| **m** | 7.516 ± 0.132 | 7.547 ± 0.154 | 7.501 ± 0.137 |
| **s** | 0.371 ± 0.087 | 0.377 ± 0.154 | 0.399 ± 0.097 |
| **ξ** | -0.250 ± 0.141 | -0.391 ± 0.389 | -0.241 ± 0.176 |
| *MSE* | 0.149 | 0.433 | 0.181 |
| 16%-percentile 84%-percentile of $M_{max}$ -estimates | 8.48 11.42 | 8.20 $3.5 \cdot 10^7$ | 8.49 14.29 |

Table B2. Comparison of the three methods of estimation for *n = 15*.

|  | *ME* | *MLE* | *PWM* |
|---|---|---|---|
| **m** | 7.516 ± 0.108 | 7.529 ± 0.117 | 7.506 ± 0.110 |
| **s** | 0.384 ± 0.076 | 0.385 ± 0.089 | 0.403 ± 0.082 |
| **ξ** | -0.239 ± 0.126 | -0.298 ± 0.259 | -0.229 ± 0.156 |
| *MSE* | 0.132 | 0.277 | 0.158 |
| 16%-percentile 84%-percentile of $M_{max}$ -estimates | 8.60 11.52 | 8.37 $4.9 \cdot 10^6$ | 8.59 14.20 |

Table B3. Comparison of the three methods of estimation for *n = 25*.

|  | **ME** | *MLE* | *PWM* |
|---|---|---|---|
| **m** | 7.511 ± 0.088 | 7.512 ± 0.092 | 7.505 ± 0.089 |
|  |  |  |  |
| **s** | 0.392 ± 0.058 | 0.390 ± 0.064 | 0.402 ± 0.062 |
| **ξ** | -0.227 ± 0.112 | -0.251 ± 0.169 | -0.219 ± 0.134 |
| *MSE* | 0.115 | 0.176 | 0.155 |
| 16%-percentile 84%-percentile of $M_{max}$ -estimates | 8.72 10.96 | 8.56 12.90 | 8.70 13.53 |

Table B4. Comparison of the three methods of estimation for *n = 50*.

| | **ME** | *MLE* | *PWM* |
|---|---|---|---|
| **m** | 7.507 ± 0.064 | 7.508 ± 0.065 | 7.503 ± 0.065 |
| **s** | 0.395 ± 0.042 | 0.392 ± 0.044 | 0.399 ± 0.044 |
| **ξ** | -0.212 ± 0.084 | -0.218 ± 0.105 | -0.205 ± 0.102 |
| *MSE* | 0.085 | 0.107 | 0.102 |
| 16%-percentile 84%-percentile of $M_{max}$ -estimates | 8.90 10.50 | 8.81 11.0 | 8.87 11.63 |

Table B5. Comparison of the three methods of estimation for *n = 200*.

| | **ME** | *MLE* | *PWM* |
|---|---|---|---|
| **m** | 7.501 ± 0.032 | 7.502 ± 0.032 | 7.501 ± 0.033 |
| **s** | 0.400 ± 0.022 | 0.400 ± 0.022 | 0.401 ± 0.023 |
| **ξ** | -0.205 ± 0.043 | -0.208 ± 0.046 | -0.203 ± 0.051 |
| *MSE* | 0.043 | 0.047 | 0.051 |
| 16%-percentile 84%-percentile of $M_{max}$ -estimates | 9.16 9.92 | 9.12 9.90 | 9.12 10.05 |

For small and moderate sample sizes *(n ≅ 10 ÷ 50)*, the *MLE* are less efficient than the *ME* and the *PWM* estimators, whereas in turn the *ME* is slightly better than the *PWM* estimator. This difference disappears for larger sample size *(n = 200)*.

There exists one additional method for the estimation of the parameters *(m,s,ξ)* suggested in (*Kijko and Graham 1998, Kijko 1999, Kijko et al. 2001*). This method is based on equating the expectation of the observed maximum magnitude $M_{max}^{obs}$ as well as the expectation of the powers $[M_{max}^{obs}]^k$, *k=2,3,...* of the observed maximum magnitude to their sample values. Thus, taking into account that $[F(x)]^n$ is the distribution function of $M_{max}^{obs}$, we get the following system of equations determining the estimates of the parameters:

$$E[M_{max}^{obs}]^k = \int_{-\infty}^{M_{max}} z^k \, d[F(z)]^n = [M_{max}]^k - k \int_{-\infty}^{M_{max}} z^{k-1} [F(z)]^n \, dz = [M_{max}^{obs}]^k, \ k=1,2,3,... \quad (B17)$$

Kijko calls the set of equation (B17) under the name *generic equations*. Since $M_{max}^{obs}$ converges in probability to the true value $M_{max}$ as *n* tends to infinity, the generic equations provide consistent estimates. But we should note that for our situation with a distribution function *GEV(x)* with a support limited from above, we have (see *Embrechts et al. 1997*)

$$(M_{max}^{obs} - M_{max}) \sim n^{\xi}, \ \ (\xi < 0), \quad (B18)$$

whereas for the moment estimators

$$(\gamma_k - g_k) \sim n^{-1/2}, \quad (B19)$$

This convergence (B19) is faster than (B18) since usually for real situations $\xi \cong -0.2 ÷ -0.3$. Thus, the generic equations provide less efficient estimators than all three types of estimators discussed above. This conclusion was confirmed by numerical examples with simulations.

**APPENDIX C: Simulation experiments on a DF for earthquake magnitudes with two branches with a maximum magnitude upper-bound.**

## C.1 Definition of the two-branch model with a maximum upper-bound

In order to estimate the biases associated with the finite sample sizes in the presence of a transition from a pure G-R law to a faster decay of the distribution of earthquake magnitudes up to an absolute upper magnitude bound $M_{max}$, it is instructive to construct an explicit model having these properties and study how extreme value theory and our methodology performs on it. In this way, we are going to be able to propose reliable corrections for systematic biases which improve significantly upon our estimations of $M_{max}$ and other characteristics of the extreme tail behavior. We thus consider the following probability density $f(x)$, consisting of two branches:

$$f(x) = \begin{cases} C\,f_1(x); & m_0 \leq x \leq m_1 \, ; \\ C\,f_2(x); & m_1 \leq x \leq M_{max}; \end{cases} \qquad \text{(C1)}$$

The density $f(x)$ is zero outside the interval $(m_0 \, ; \, M_{max})$, $C$ is a normalizing constant.
The branches should satisfy the following properties:

(i)    The first branch $f_1(x)$ should be the pure Gutenberg-Richer law;
(ii)   The second branch $f_2(x)$ should decrease to zero at some finite upper point;
(iii)  At the point $m_1$ of connection of the two branches, the density $f(x)$ and its first derivative should be continuous.

The density $f(x)$ defined in (C1) has the shape of a "duck beak": it may extend to a quite distant end point $M_{max}$ at a very low level, so that events with magnitudes close to $M_{max}$ may occur very, very seldom.
In accordance with the first condition, we write

$$f_1(x) = \beta \, exp(\, -\beta(x - m_0)\,); \qquad \beta > 0; \quad m_0 \leq x \leq m_1 \, . \qquad \text{(C2)}$$

We have decided to parameterize the second branch in the following form:

$$f_2(x) = C_1(1/\sigma)[1 - \alpha(x - m_1)/\sigma]^{1/\alpha - 1} \, ; \qquad \alpha > 0; \quad m_1 \leq x \leq M_{max} = m_1 + \sigma/\alpha \, , \qquad \text{(C3)}$$

where $C_1$ is a constant. We consider the parameters $\beta$, $m_0$ as known, whereas the parameters $C_1$, $\alpha$, $\sigma$, $m_1$ (or $M_{max}$) are free for fitting. The third condition (continuity of density and its first derivative) introduces two equations for these four parameters. Thus, we are left with only two parameters for fitting. We can choose any two from these four parameters for fitting, e.g. $m_1$, $M_{max}$. In this case, the other parameters are expressed as a function of $\beta$, $m_0$, $m_1$, $M_{max}$ as

$$\alpha = 1/(1 + \beta(\, M_{max} - m_1);$$

$$\sigma = (\, M_{max} - m_1)/\, (1 + \beta(\, M_{max} - m_1); \qquad \text{(C4)}$$

$$C_1 = \beta\sigma \, exp(-\beta(\, m_1 - m_0)).$$

## C.2 Fit of the two-branch model to the Harvard catalog

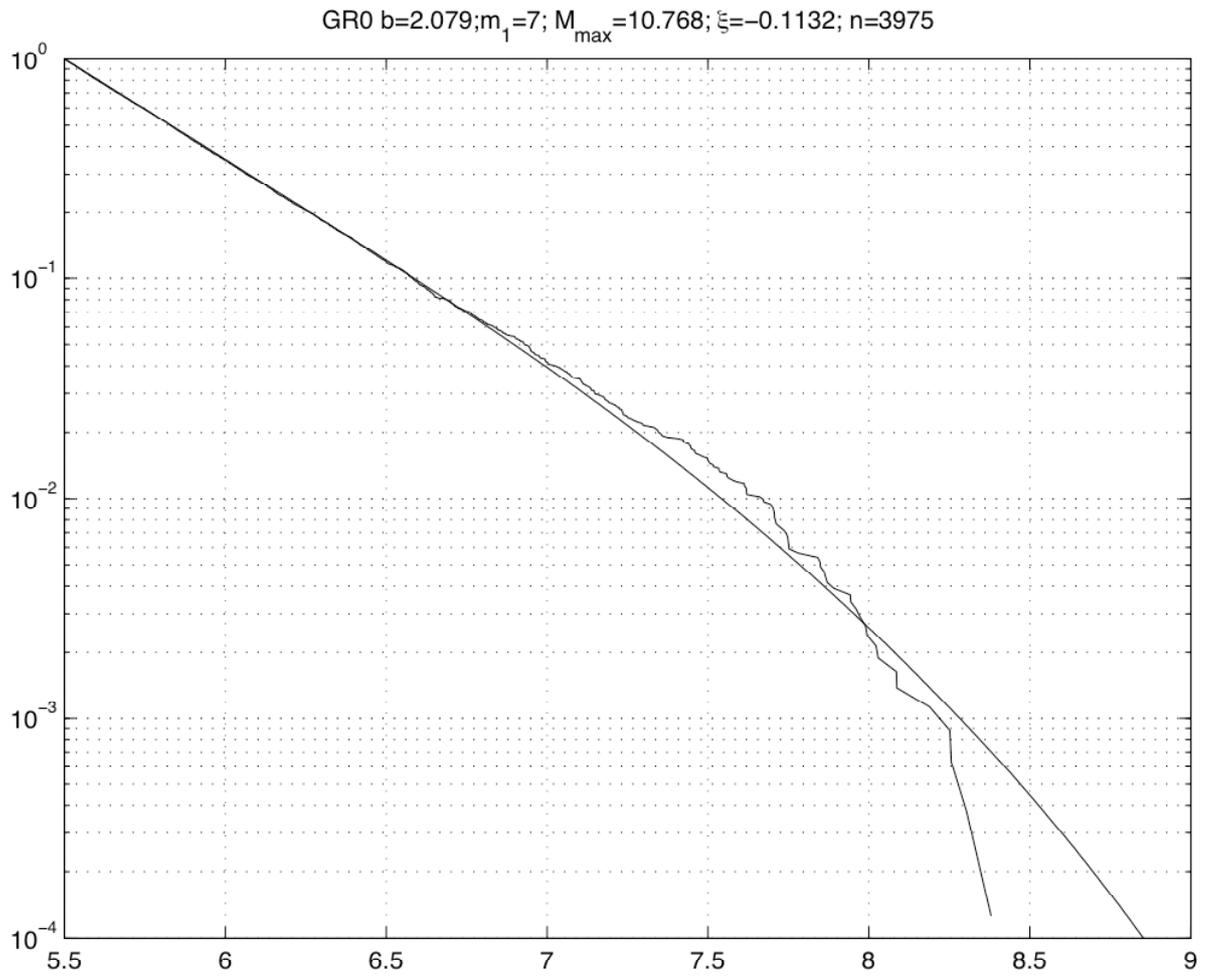

Fig.C1: One among several almost equivalent fits (with practically the same likelihood values) of the global Harvard catalog (*01.01.77-18.12.04; $m_W \geq 5.5$; $h \leq 70$ km*; main shocks, *N=3975*) with the two-branch model (C1-C3). The parameters are $\beta=2.079$, $m_1=7.0$, $M_{max}=10.768$, $\xi=-01132$.

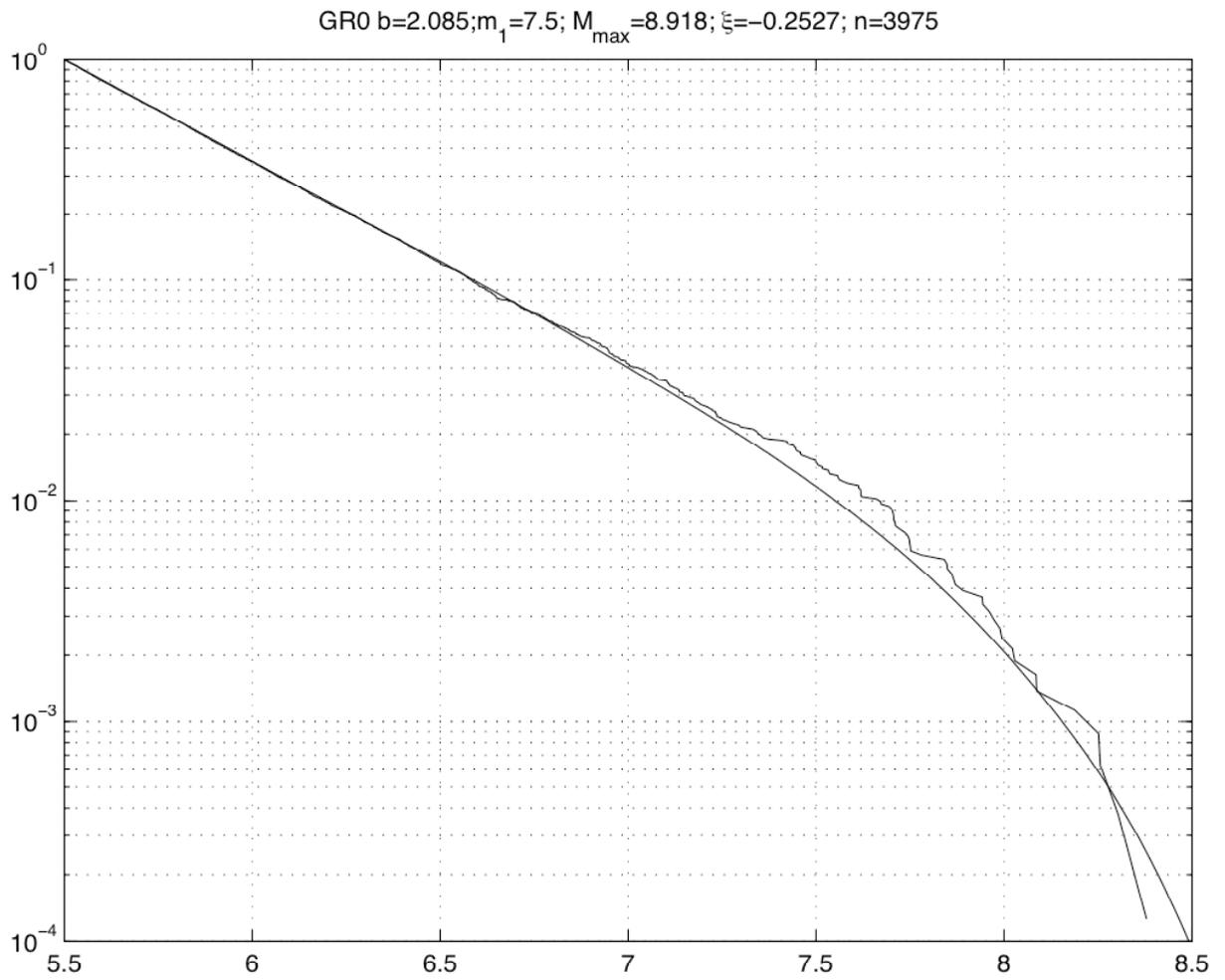

Fig.C2: Same as Fig.C1. The parameters are now $\beta=2.085,\ m_1=7.5,\ M_{max}=8.918,\ \xi=-0.2527$.

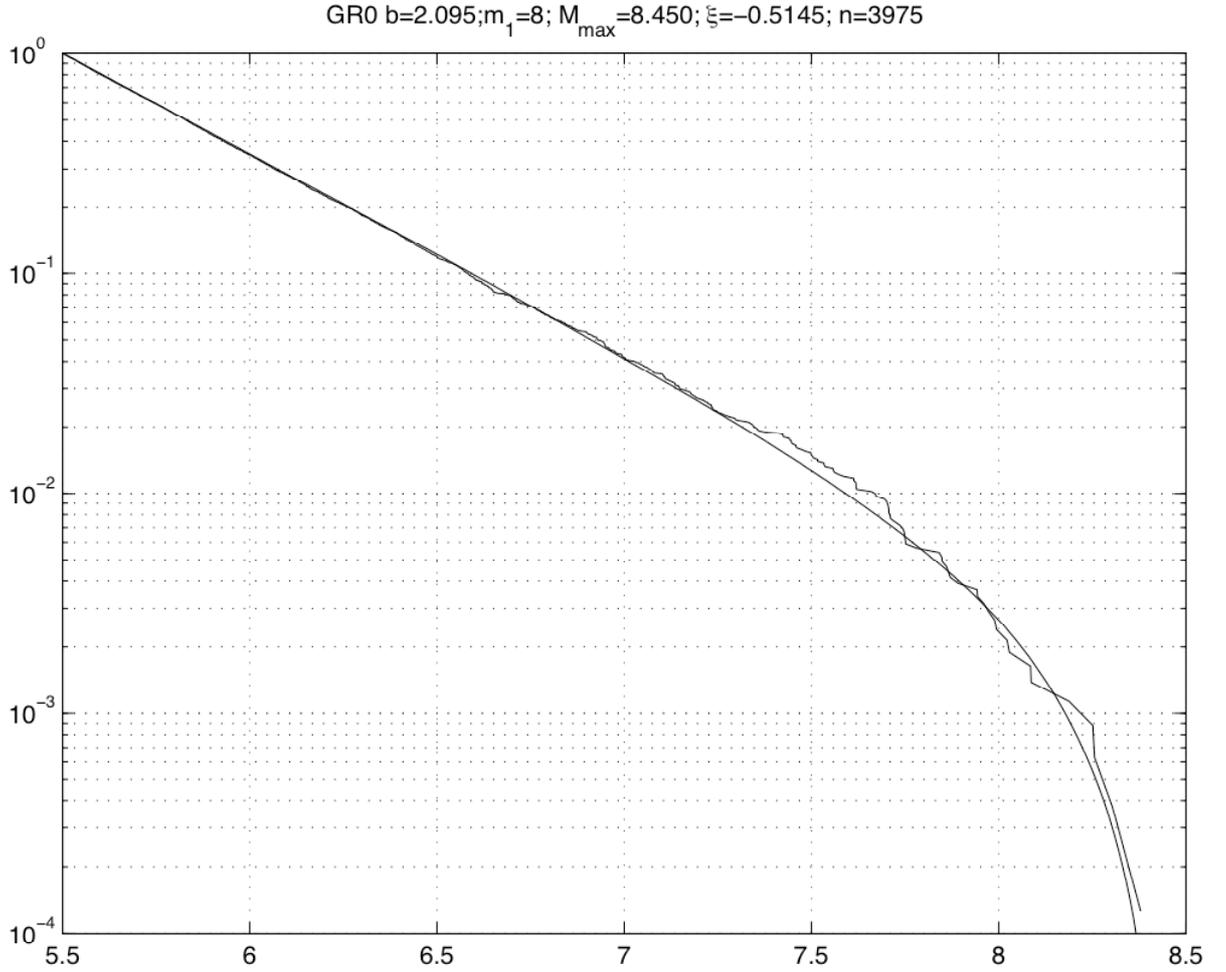

Fig.C3: Same as Fig.C1. The parameters are now *β=2.095, m₁=8.0, M_max=8.450, ξ=-0.5145*.

Let us illustrate the relevance of our two-branch model (C1-C3) by exploring its ability to fit the global Harvard catalog (*01.01.77-18.12.04; m_W ≥ 5.5; h ≤ 70 km*; main shocks, *N=3975*). The Figures C1-C3 show how three different pairs of the parameters *(m₁, M_max)* can approximate the sample tail *1 − F(x)* of the catalog. All these pairs give very close likelihood values, so none of them has an appreciable advantage over the others in terms of the statistical likelihood. One can observe that these three fits look rather satisfactorily in spite of the fact that the corresponding parameters differ considerably. This means that the estimation of the two-branch model (C1-C3) is rather uncertain. Unsurprisingly, the problem comes from the fact that the data sample has relatively few observations *N(m₁)* exceeding *m₁* (which provide a direct constraint in the tail behavior):

$$N(m_{1=7.0}) = 166 \ (Fig.C1); \quad N(m_{1=7.5}) = 59 \ (Fig.C2); \ N(m_{1=8.0}) = 9 \ (Fig.C3) \qquad (C5)$$

This means that this uncertainty and the degenerate nature of the fit is inherent.

### C.3 Degenerate behavior of the estimates of the parameters *(m₁, M_max)*.

It is necessary to discuss an unpleasant but unavoidable property of the estimates of the parameters *(m₁, M_max)* of the two-branch model (C1-C3): if, for a fixed sample size *N*, the true *m₁* is too close to the true *M_max*, then the Moment-estimates of these parameters merge with a large probability, i.e. the difference between these estimates become very small. As a consequence, one is not able to detect the second branch which reduces to the upper bound $M_{max}$. If one considers samples generated with a fixed difference $(M_{max} - m_1)$, small sample sizes $N$ will not be able to separate between $m_1$ and $M_{max}$ but, by increasing $N$, we can always reach a sufficiently large $N$ such that $M_{max}$ and $m_1$ can be separated by the ME method. It turns out that for $n=3975$ (sample size of the main shocks in the global Harvard catalog, 1977-2004), the ME method is able to distinguish with good reliability the differences $(M_{max} - m_1) \cong 1.5$-$2$ and larger.

We have performed extensive simulations for different sample sizes, $N=1300$ (number of main shocks at shallow depths in subduction zones, 1977-2000), $N=4000, N=10000, N=20000$ for different $Mmax = 9.5; 10.0; 10.5; 11.0$, and for $6.5 \leq m_1 < M_{max}$. These simulations show that the probability $P\{ M_{max} - max(x_k) < 0.1\}$ for the difference $M_{max} - max(x_k)$ to be small is significant (larger than 0.1) for

$$
\begin{aligned}
&m_1 > 6.5 \quad (N=1300); \\
&m_1 > 7.25 \quad (N=4000); \\
&m_1 > 7.5 \quad (N=10000); \\
&m_1 > 8.0 \quad (N=20000).
\end{aligned}
\qquad (C9)
$$

Looking at the empirical sample tails in Figs. C1-C3, we could conclude visually (of course it is not a strict conclusion) that the magnitude (an estimate of $m_1$) beyond which the curvature corresponding to the second branch of the distribution given by equations (C1-C3) becomes significant can hardly be less than 7.0. Reading from the set of inequalities (C9), we would conclude that the estimation of $m_1$ for a sample of $N=1287$ events as in the subduction zones is practically hopeless, whereas there remains still some hope to estimate $m_1$ for the global catalog with $N=3975$ main shocks. The average number Av#$(m_W > m_1)$ of observations exceeding $m_1$ which are needed to make $P\{ M_{max} - max(x_k) < 0.1\}$ small (less than 0.1) is found approximately to be Av#$(m_W > m_1) > \sim 100$. For the empirical global Harvard catalog, the 100-th order statistics is $m_W = 7.23$ which is slightly less that $m_W = 7.25$ shown in the set of inequalities (C9) as the approximate threshold that needs to be passed in order to distinguish between $M_{max}$ and $m_1$ for $N=4000$. When Av#$(m_W > m_1) > < 100$ or equivalently when the inequalities (C9) are obeyed, we observe considerable biases in the median and quantile estimates of $M_{max}$ and of $m_1$.

The overall conclusion of these tests is that the direct estimation of the parameters of the two-branch model (C1-C3) using the catalogue of magnitudes is questionable even for sample sizes $N=4000$ comparable with the global Harvard catalog, and becomes hopeless for the smaller data size $N=1300$ corresponding to the main shocks at shallow depths in subduction zones. Similar limitations are observed for other parameterization *(Pisarenko and Sornette, 2003; 2004)*. We are therefore forced to use another approach to estimate the parameters describing the tail of the distribution of earthquake magnitudes. It turns out that working with the sample of $n$ $T$-maxima of magnitudes in windows of size $T$ provides more stable and reliable estimates. The two-branch model (C1-C3) will be used further only for estimating the scatter of the statistical characteristics (std and quantiles) of the estimates of empirical samples by means of synthetic simulations, and not for the direct estimation of the tail parameters.

**References.**

Bender, B.K. and D.M. Perkins (1993) Treatment of parameter uncertainty and variability for a single seismic hazard map, Earthquake Spectra 9 (2), 165-195.

Bird, P., and Y. Y. Kagan (2004) Plate-tectonic analysis of shallow seismicity: Apparent boundary width, beta, corner magnitude, coupled lithosphere thickness, and coupling in seven tectonic settings, Bull. Seismol. Soc. Am., 94 (6), 2380-2399.

Black, N., Jackson, D. and Rockwell, T. (2004) Maximum Magnitude in Relation to Mapped Fault Length and Fault Rupture, American Geophysical Union, Fall Meeting 2004, abstract #S41A-0922.

Christopeit N. (1994) Estimating parameters of an extreme value distribution by methods of moments, Journal of Statistical Planning and Inference, 41, 173-186.

Coles S. and M.Dixon (1999) Likelihood-based Inference for Extreme Value Models, Extremes, 2:1, 5-23.

Cosentino P., Ficara V., and D.Luzio (1977) Truncated exponential frequency-magnitude relationship in the earthquake statistics. Bull. Seism. Soc. Am. 67, 1615-1623.

Cox D.R. and P.A. Lewis (1966) The Statistical Analysis of Series of Events, John Wiley, London-N.-Y.).

Dargahi-Noubary G.R. (1983) A procedure for estimation of the upper bound for earthquake magnitudes. Phys. Earth Planet Interiors. 33, 91-93.

Epstein, B.C. and C. Lomnitz (1966) A model for the occurrence of large earthquakes, Nature 211, 954-956.

Gutenberg, B. and C. F. Richter (1942) Earthquake magnitude, intensity, energy and acceleration, *Bull. Seism. Soc. Am.* 32, 163-191.

Gutenberg B., Richter C. (1954) Seismicity of the Earth. 2-nd Edition, Princeton Univ. Press.

Gutenberg, B. and C. Richter (1956)  Earthquake magnitude, intensity, energy, and acceleration, part II, *Bull. Seism. Soc. Am.* 46, 105-145.

Hosking, J.R., Wallis, J.R., and Wood, E.F. (1985) Estimation of the Generalized Extreme-Value Distribution by the Method of Probability-Weighted Moments, Technometrics 27, 251-261.

Kagan Y.Y. (1991) Seismic moment distribution. Geophys. J. Int. 106, 123-134.

Kagan Y.Y. (1996) Comment on "The Gutenberg-Richter or characteristic earthquake distribution, which is it?" by Steven G. Wesnousky. Bull. Seism. Soc. Am. 86, 274-285.

Kagan Y.Y. (1997) Seismic moment-frequency relation for shallow earthquakes: Regional comparison. Journ. of Geophys. Research. 102, 2835-2852.

Kagan Y.Y. (1997) Earthquake Size Distribution and Earthquake Insurance. Commun. Statist.-Stachastic Models, 13(4), 775-797.

Kagan Y.Y. (1999) Universality of the Seismic Moment-frequency Relation. Pure Appl. Geophys. 155, 537-573.

Kagan Y.Y. (2002) Seismic moment distribution revisited: I. Statistical results. Geophys. J. Int. 148, 520-541.

Kagan Y.Y. (2002) Seismic moment distribution revisited: II. Moment conservation principle. Geophys. J. Int. 149, 731-754.

Kijko A. Graham G. (1998) Parametric-historic Procedure for Probabilistic Seismic Hazard Analysis, Pure and Applied Geophysics, 152, 413-442.

Kijko A. (1999) Statistical Estimation of Maximum Regional Earthquake Magnitude $m_{max}$ , 12-th European Conference on Earthquake Engineering, FW:022, 1-22.

Kijko A., Lasocki S. and G.Graham (2001) Non-parametric Seismic Hazard in Mines, Pure and Applied Geophysics, 158, 1655-1675.

Knopoff L. and Kagan Y. (1977) Analysis of the Extremes as Applied to Earthquake Problems, J. Geophys. Res. 82, 5647–5657.

Knopoff L., Kagan Y., and R. Knopoff (1982) b-values for foreshocks and aftershocks in real and simulated earthquake sequences, Bull. Seism. Soc. Amer.72 (5), 1663-1675.

Kagan Y.Y. and F. Schoenberg (2001) Estimation of the upper cutoff parameter for the tapered distribution. J. Appl. Probab. 38A, 901-918.

Kijko, A. and G. Graham (1998) Parametric-historic procedure for probabilistic seismic hazard analysis. Part I: estimation of maximum regional magnitude Mmax, Pure Appl. Geophys. 152, 413-442.

Kijko A. and Sellevol M.A. (1989) Estimation of earthquake hazard parameters from incomplete data files. Part I, Utilization of extreme and complete catalogues with different threshold magnitudes. Bull. Seism. Soc. Am. 79, 645-654.

Kijko A. and Sellevol M.A. (1992) Estimation of earthquake hazard parameters from incomplete data files. Part II, Incorporation of magnitude heterogeneity. Bull. Seism. Soc. Am. 82, 120-134.

Kijko, A., S. Lasocki and G. Graham (2001) Non-parametric seismic hazard in mines. Pure Appl. Geophys. 158, 1655-1675.

Main Y., Irving D., Musson R. (1999) and A.Reading. Constraints on frequency-magnitude relation and maximum magnitudes in the UK from observed seismicity and glacio-isostatic recovery rates. Geophys. J. Int. 137, 535-550.

Main Y. (2000) Apparent Breaks in Scaling in Earthquake Cumulative Frequency-Magnitude Distribution: Fact or Artifact? Bull. Seism. Soc. Am. 90, 86-97.

Molchan G., Kronrod T., and Panza G.F. (1997) Multi-scale seismicity model for seismic risk. Bull. Seis. Soc. Am. 87, 1220-1229.

Ogata Y., and K. Katsura (1993) Analysis of temporal and special heterogeneity of magnitude frequency distribution inferred from earthquake catalogues. Geophys. J. Int. 113, 727-738.

Pisarenko V.F. (1991) Statistical evaluation of maximum possible magnitude. Izvestia, Earth Physics, 27, 757-763.

Pisarenko V.F., Lyubushin A.A., Lysenko V.B., and T.V.Golubeva (1996) Statistical Estimation of Seismic Hazard Parameters: maximum possible magnitude and related parameters. Bull. Seism. Soc. Am. 86, 691-700.

Pisarenko V.F., and D.Sornette (2003) Characterization of Frequency of Extreme Earthquake Events by the Generalized Pareto Distribution. Pure Appl. Geophys. 160, 2343-2364.

Pisarenko V.F., and D.Sornette (2004) Statistical Detection and Characterization of a Deviation from the Gutenberg-Richter Distribution above Magnitude 8. Pure Appl. Geophys. 161, 839-864.

Smith R. (1985) Maximum Likelihood Estimation in a Class of Non-Regular Cases, Biometrika, 72, 67-92.

Sornette, A. and Sornette, D. (1989) Self-organized criticality and earthquakes, Europhys. Lett. 9, 197-202.

Sornette, A. and Sornette, D. (1990) Earthquake rupture as a critical point, Tectonophysics, 179, 327-334.

Sornette, A., Davy, P. and Sornette, D. (1990) Growth of fractal fault patterns, Phys. Rev. Letters, 65, 2266-22-69.

Sornette, A. and Sornette, D. (1999) Renormalization of earthquake aftershocks, Geophys. Res. Lett., 26, 1981-1984.

Sornette, D., L. Knopoff, Y.Y. Kagan and C. Vanneste (1996) Rank-ordering statistics of extreme events : application to the distribution of large earthquakes, J.Geophys.Res. 101, 13883-13893.

Utsu T. (1999) Representation and Analysis of the Earthquake Size Distribution: A Historical Review and Some New Approaches. Pure Appl. Geophys. 155, 509-535.

Ward, S.N. (1997) More on Mmax, Bulletin of the Seismological Society of America 87 (5), 1199-1208.

Wesnousky S.G. (1994) The Gutenberg-Richter or characteristic earthquake distribution, which is it? Bull. Seism. Soc. Am. 84, 1940-1959.

Wu Z.L. (2000) Frequency-size distribution of global seismicity seen from broad-band radiated energy. Geophys. J. Int. 142, 59-66.